\documentclass[12pt,a4paper]{article}
\usepackage[utf8]{inputenc}
\usepackage[english]{babel}
\usepackage{graphicx}
\usepackage{comment}
\usepackage{authblk}

\usepackage[margin=0.75in]{geometry}
\usepackage{fullpage}

\usepackage{times}
\usepackage{fancyhdr,graphicx,amsmath,amssymb}
\usepackage[ruled,vlined]{algorithm2e}
\include{pythonlisting}

\usepackage{tikz-cd}
\usepackage{amssymb}
\usepackage{amsmath}
\usepackage{amsthm}
\usepackage[acronym]
{glossaries}
\usepackage{times}

\usepackage{nomencl}
\makenomenclature

\newtheorem{teo}{Theorem}[section]
\newtheorem{defi}[teo]{Definition}

\makeglossaries

\usepackage{graphicx}
\graphicspath{ {.pdf} }
\usepackage{wrapfig}
\usepackage{subcaption}

\begin{document}

\date{}

\title{Construction of edge-ordered multidirected graphlets for comparing dynamics of spatial temporal neural networks}
\author[1,3]{Joshua M. Roldan}
\author[1,3]{Sebastian Pardo G.}
\author[1,3]{Vivek Kurien George}
\author[1,2,3]{Gabriel A. Silva \footnote{Corresponding author: G.S. Email: gsilva@ucsd.edu}}
\affil[1]{Department of Bioengineering, University of California San Diego}
\affil[2]{Department of Neurosciences, University of California San Diego}
\affil[3]{Center for Engineered Natural Intelligence, University of California San Diego}

\renewcommand\Affilfont{\small}
\maketitle

\renewenvironment{abstract}
{\begin{quote}
\noindent \rule{\linewidth}{.5pt}\par{\bfseries \abstractname.}}
{\medskip\noindent \rule{\linewidth}{.5pt}
\end{quote}
}

\begin{abstract}
The integration and transmission of information in the brain are dependent on the interplay between structural and dynamical properties. Implicit in any pursuit aimed at understanding neural dynamics from appropriate sets of mathematically bounded conditions is the notion of an underlying fundamental structure-function constraint imposed by the geometry of the structural networks and the resultant latencies involved with transfer of information. We recently described the construction and theoretical analysis of a framework that models how local structure-function rules give rise to emergent global dynamics on a neural network. An important part of this research program is the requirement for a set of mathematical methods that allow us to catalog, theoretically analyze, and numerically study the rich  dynamical patterns that result. One direction we are exploring is an extension of the theory of graphlets. In this paper we introduce an extension of graphlets and associated metric that maps the topological transition of a network from one moment in time to another at the same time that causal relationships are preserved. \\
\end{abstract}

\section{Introduction}
The integration and transmission of information in the brain are dependent on the interplay between  structural and dynamical properties across many scales of organization. This interplay happens in molecular and diffusion interactions occurring in neurons and other neural cells, and at physiological scales in individual single cells, networks of cells, and eventually in networks of brain regions. A critical consideration towards a systems engineering view of the brain is understanding how mathematically imposed constraints resulting from physical considerations determine neural dynamics and what the brain is able to do.

Implicit in any theoretical or computational pursuit aimed at understanding neural dynamics from appropriate sets of mathematically bounded conditions, whether acknowledged or not, is the notion of an underlying fundamental structure-function constraint. This fundamental constraint is imposed by the geometry of the structural networks that make up the brain at different scales, and the resultant latencies involved with the flow and transfer of information within and between functional scales. It is a constraint produced by the very way the brain is wired up, and how its constituent parts necessarily interact, e.g. neurons at one scale and brain regions at a higher scale. The networks that make up the brain, across all the various scales of organization, are physical constructions over which signals and information must travel. These signals are subject to processing times and signaling speeds (conduction velocities) that must travel finite distances to exert their effects - to transfer the information they are contributing to the next stage in the system. Nothing is infinitely fast. Furthermore, the latencies created by the interplay between structural geometries and signaling speeds is generally at a temporal scale similar to the functional processes being considered. So it matters from the perspective of understanding how structure determines function in the brain, and how function modulates structure, for example, in learning or plasticity. 

Towards such a systems engineering understanding of the brain, on-going efforts in our group are focused on deriving and studying the mathematical relationships and consequences that this fundamental structure-function constraint produces. Central to the thesis of our work is the notion that the mathematical relationships we discover and prove about information integration and computation in biological neural networks should be independent, as much as possible, from unnecessary physiological details. Our objective is to identify and understand the most basic and simple set of conditions that necessarily result, and to write them down in mathematical forms amenable to deep theoretical analyses. We are attempting to discover the fundamental algorithms associated with neural dynamics and neurobiological information representations. By 'unnecessary biological and physiological details' we mean that our theoretical constructions should retain only features deemed essential to the algorithms themselves, while remaining as much as possible independent of details responsible for their implementation in the 'wetware' environment of the brain. The algorithms and mathematical relationships that underly them should be independent of the neurobiological specifics of any particular experimental model, for example. 

We recently described the construction and theoretical analysis of a framework derived from the canonical neurophysiological principles of spatial and temporal summation. This framework models the competing interactions of signals incident on a target downstream node (e.g. a neuron) along directed edges coming from other upstream nodes that connect into it in a network \cite{Silva 2, Silva}. We considered how temporal latencies produce offsets in the timing of the summation of incoming discrete events due to the geometry (physical structure) of the network, and how this results in the activation of the target node. The framework we constructed models how the timing of different signals compete to ‘activate’ nodes they connect into. This could be a network of neurons or a network of brain regions, for example. At the core of the model is the notion of a refractory period or refractory state for each node. This reflects a period of internal processing, or period of inability to react to other inputs at the individual node level. It is important to note that we did not assume anything about the internal model that produces this refractory state, which could include an internal processing time during which the node is making a decision about how to react. In a geometric network temporal latencies are due to the relationship between signaling speeds (conduction velocities) and the geometry of the edges on the network (i.e. edge path lengths). We have shown that the interplay between temporal latencies of propagating discrete signaling events on the network relative to the internal dynamics of the individual nodes - when they become refractory and for how long - can have profound effects on the dynamics of the network \cite{Silva 2}, and the definition of mathematical bounds on conditions required for a formal definition of efficient signaling \cite{Silva}. We have also shown that Basket cell neurons optimize their morphology in order to preserve what we call the refraction ratio, a balance between the temporal dynamics of individual nodes relative to the dynamics of the entire network \cite{Puppo}. Our model allows us to compute and study the local dynamics that govern and give rise to emergent global dynamics on a network. This framework and its theoretical analysis are a concrete example of brain invoked algorithms that directly result from the fundamental structure-function constraint, and are independent of any neurobiological or biophysical implementation details. We refer the reader to the relevant references for a full discussion \cite{Silva}. 

An important part of this research program is the requirement for a set of mathematical methods that allow us to catalog, theoretically analyze, and numerically study the rich dynamic patterns that result from network simulations using our framework. One direction we are exploring is an extension of the theory of graphlets, and the technical focus of the work in this paper. Our motivation and use of these results is their application to neural dynamics broadly speaking and the analysis of numerical simulations from our framework applied to the study of the brain and to the development of a new machine learning architecture we are constructing. But the work itself if of even broader context and utility, with applications to network theory and the analyses of a wide range of dynamic spatial-temporal networks, and to mathematics in the sense that we are extending the theory of graphlets. Specifically, we introduce an extension of graphlets that maps the topological transition of a network from one moment in time to another at the same time that causal relationships are preserved, including in situations where multiple signals contribute to the initiation of a downstream activation event in a target node.

The remainder of this paper is organized as follows: Section \ref{sec:background} discusses some background and other related work. It provides a  short summary of previous work dealing with network motifs and graphlets. Additional information is provided in the Appendix. In Section \ref{sec:prelim} we give a brief introduction to different types of networks and how graphlets are derived. Section \ref{sec:signalgraphs} constructs synaptic signal graphs and introduces graphlets for multidigraphs and edge-ordered multidigraphs. In Section \ref{sec:stats} we describe an extension of the graphlet-orbit transition metric to edge-ordered multidigraphs. Section \ref{sec:conclusions} discusses some concluding remarks.

\section{Background and Prior Relevant Work} \label{sec:background}
Milo \emph{et. al.} were the first to observe that certain subgraphs, called motifs, occurred in networks with a much higher frequency than in similar randomized networks \cite{Milo 1}. Randomized networks preserved the same number of edges coming in and out as the corresponding node from the real network, but were assigned random connectivity. Motifs were considered the building blocks of larger networks. Network motifs characterized classes of networks whose properties are defined by specific types of elementary structures. Examples of networks motifs include \textit{feed- forward loops} and \textit{Bi-fans} found in \textit{transcription networks} and \textit{neuron synaptic connection networks}, respectively (also see \cite{Milo 1}). 

In 2004, Pr\v{z}ulj, introduced the concept of \textit{graphlets}, which are small connected induced subgraphs of a larger network. Graphlets are similar to network motifs in the sense that both analyze networks using subgraphs. However, graphlets are induced subgraphs, whereas network motifs allow partial subgraphs. Graphlets can be understood as a generalization of node degrees. Instead of just looking at how many other nodes a given node touches, one asks how many triangles, chains, or stars structures does a node participate in. This notion better allows a better answer to the question 'how many times does each subgraph show up in the graph?', referred to as the the \textit{Subgraph Counting Problem}. Compared to network motifs, graphlets have an advantage in that they do not depend on generating corresponding randomized networks, and subgraphs are induced. Different randomizations can produce different network motifs, so consistency is an issue. Requiring graphlets be induced subgraphs rather than partial subgraphs yields increased network measure precisions. 

Graphlets have found merit in their application to Protein-Protein Interaction (PPI) networks (see \cite{Prz 1}). In 2007 Pr\v{z}ulj introduced a similarity metric called the \textit{graphlet degree distribution agreement} (see \cite{Prz 2}) which showed to be more efficient in comparing two static networks than network motifs or other existing measures. Graphlets were first introduced for undirected networks and then extended to directed networks (see \cite{Prz 3}); the latter captured more information and increased the precision by improving the degree of granularity used to describe a network. While we wanted to use graphlets to catalog and study causal signal patterns that resulted from our model, the signal dynamics we need to analyze are intrinsically temporal. Thus, what we need is an inherently temporal or dynamic graphlet construction and associated metric  in order to compare different patterns.

Milenkovi\'c et al. made a first attempt to transform graphlets into a temporal framework in hopes of capturing more information than the original static version of the theory. In \cite{Holme}, Holme and Saram\"{a}ki defined a temporal network as a set of nodes and a set of \textit{events} (temporal edges) that are associated with two time parameters: a start time and a duration time. Any temporal network can be modeled as a \textit{sequence of snapshots}. Each of these snapshots is a static graph which aggregates the temporal information observed within a fixed time interval. The new temporal version of graphlets was called \textit{dynamic graphlets} by Milenkovi\'c and colleagues (see \cite{Milen}). These are equivalence classes of isomorphic time-connected temporal subgraphs. More precisely, two induced temporal subgraphs are equivalent if they are structurally isomorphic and have the same relative order of events. Dynamic graphlets were shown to be superior relative to static and static-temporal approaches in their ability to perform network classification and node classification. The additional contribution of dynamic graphlets is the acquired ability to capture relationships between snapshots.

But although dynamic graphlets are able to produce greater accuracy for temporal network classification than earlier methods, there persist a number of fundamental limitations. Dynamic graphlets do not inform how the local topology changes over time, e.g., where poorly connected subgraphs become near-cliques. This notion of measuring how the local topology of a network changes over time was the motivation behind work by Ribiero, Silva and Aparicio who proposed a complementary graphlet-based comparison metric of temporal networks by defining \textit{graphlet-orbit transitions}. Graphlet-orbit transitions account for the possible change of states of a node appearing in different orbits (see  \cite{Temp Net}); e.g., a node in a periphery of a star moves to the center. 

For our purposes though, the framework needs to be extended further. Dynamic graphlets only allow one event or temporal edge configuring into paths. This makes it incapable of dealing with networks that have multiple (signaling) events or multiple edges. But this is critical to our model because multiple signals from the same upstream node activated repeatedly can contribute to the same running summation that triggers a single activation event in the downstream node it connects into. This implies that we need to represent multiple edges between connected nodes pairs, which dynamic graphlets cannot do. To address this, we extended graphlet-orbit transitions to incorporate directed edges and multi-directed edges. 
 
 The relationship between dynamic graphlets and graphlet-orbit transitions is complementary. Dynamic graphlets are essentially causal paths obtained from jumping from one snapshot to the next. Thus, dynamic graphlets inherently preserve causal order. On the other hand, graphlet-orbit transitions capture information within a snapshot and its transition to the next one, but it blurs the more granular causal information between snapshots, including the order in which the edges that encode signals arrive due to the aggregation of the dynamics. Here, we take these two graphlet-based temporal metrics and combine them into one framework, so that the topological transition from one snapshot to another can be characterized at the same time that causal relationships are preserved, including in situations where multiple signals contribute to the initiation of a single downstream activation event.

\section{Preliminaries} \label{sec:prelim}
In this section we summarize the principle ideas and notation of graphlets and put them in context with standard graph theory. The mathematical notation was chosen to be as consistent as possible with the existing graphlet literature,and the theoretical construction of our neuro-derived dynamic signaling framework.

\subsection{Undirected graphs}
Graphs consist of vertices (or nodes - we use these terms interchangeably here) and edges, where vertices represent the components of a system and edges the relationships between these components. Formally, a \textit{graph} $G(V,E)$ is defined by a set of vertices $V(G)$, and a set of edges $E(G)$ such that $E(G)\subseteq V(G)\times V(G)$.

\begin{figure}[h]
    \centering 
    \includegraphics [width=1.5in]{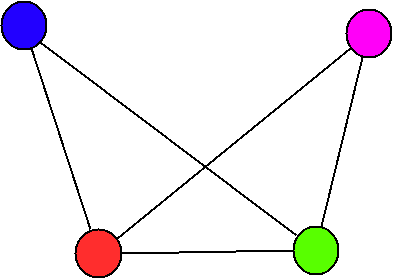}
    \caption{A graph $G$ with four vertices.}
   \label{fig:Ex graph 1}
\end{figure}

\begin{figure}[h!] \label{fig:subgraphs}
    \centering 
    \includegraphics [width=3in]{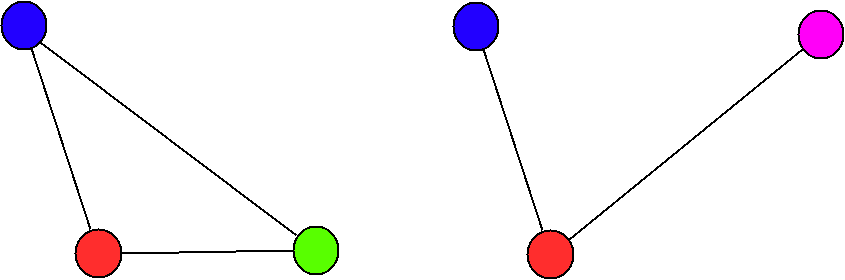}
    \caption{Two subgraphs of $G$ from Figure \ref{fig:Ex graph 1} of size 3.}
    \label{fig:Ex graph 2}
\end{figure}

One can always consider subsets $V'(G)$ $\subseteq$ $V(G)$ and $E'(G)$ $\subseteq$ $E(G)$. Figure \ref{fig:subgraphs} illustrates two   \textit{subgraphs} of $G$.

The size of a graph $G$ is defined as the total number of elements in the vertex set $|V(G)|$. Since any subgraph is a graph by itself, we also say that the size of a subgraph $H$ is $n$ if $|V(H)|$ = $n$. We denote this by $H(n)$. For example, the graph $G$ in Figure \ref{fig:Ex graph 1} has size $4$ and both subgraphs are of size $n=3$, Figure \ref{fig:Ex graph 2}.

\begin{figure}[h!]
    \centering 
    \includegraphics [width=4.5in]{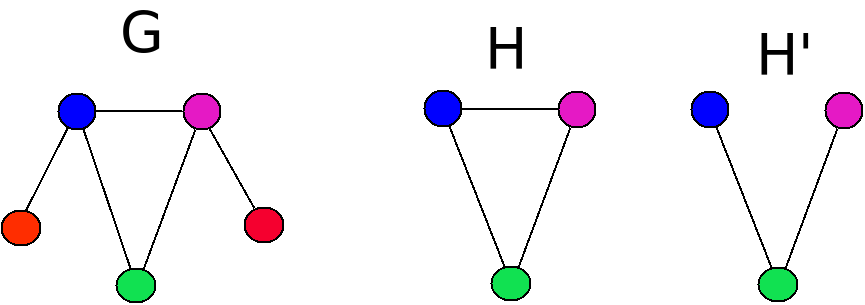}
    \caption{Give the graph $G$, $H$ is an induced subgraph of $G$ constructed by vertex deletion of the two red vertices. However, $H'$ is not induced because it additionally contains an edge deletion not induced by vertex deletion.}
   \label{fig:Ex graph 3}
\end{figure}

A special kind of subgraph are induced subgraphs. Induced subgraphs are obtained by vertex deletion (as opposed to edge deleted subgraphs constructed by deleting edges but preserving vertices). Given a graph $G$, let $X$ be a set of vertices deleted from $G$. The resulting subgraph $G - X$ is an induced subgraph. Note that we are actually interested in the subgraph constructed (i.e. induced) by the set of vertices that are \emph{not} deleted, not the set of vertices that are. Explicitly, $G[Y] =  G - X$ is the subgraph induced by the vertices in the subgraph $[Y]$. The edge set of $G[Y]$ are all edges in $G$ that have both endpoints in $G[Y]$. For example, given $G$ in Figure \ref{fig:Ex graph 3}, $H$ is an induced subgraph of $G$ constructed by vertex deletion of the two red vertices, but $H'$ is not induced because it additionally contains an edge deletion not induced by vertex deletion. They are both however, subgraphs of $G$.

In mathematics, functions are used to relate two objects having the same structure in order to transfer information. In the graph context, these functions have the feature of preserving the connections between nodes, also called adjacencies. Given two graphs $G$ and $G'$, a \textit{graph homomorphism} $f:G\longrightarrow G'$ is a function sending the vertex set $V(G)$ to the vertex set $V(G')$ in such a way that adjacencies are preserved. That is, if $(u,v)$ $\in$ $E(G)$, then $(f(u),f(v))$ $\in$ $E(G')$. When the function $f$ is also a bijective function, we say that $f$ is a graph \textit{isomorphism}, wherewith $G$ and $G'$ are said to be \textit{isomorphic} graphs. Intuitively, two graphs $G$ and $G'$ are isomorphic if we can go from $G$ to $G'$ (or vice versa) just by relabeling the vertices without changing their topology. 

A special case of isomorphism arises when the function $f$ has the same domain and range, that is, when $f$ is an isomorphism from $G$ to itself. These particular functions are called \textit{automorphisms}. Automorphisms play a key role in classifying the local topological properties of a vertex. The set of all automorphism under composition defines a group called the \textit{automorphism group} of $G$ which we denoted by $Auto(G)$ (see \cite{Chart}). The automorphism group $Auto(G)$ induces a partition of $G$ into equivalence classes where two vertices $u$ and $v$ are equivalent if there exists an automorphism $f$ such that $f(u) = v$. Vertices of the same equivalence class are said to be in the same \textit{automorphism orbit} or \textit{orbit} for short. Thus, two vertices in a graph are topologically identical if and only if they belong to the same automorphism orbit. A similar concept is edge automorphism orbits. One can utilize small connected non-isomorphic induced subgraphs called \textit{graphlets} and their corresponding orbits to characterize the vertices of a larger graph, and hence the larger graph itself. Our work in this paper focuses on vertex graphlets and leaves edge-graphlets for future work.

From a theoretical perspective, the most informative way of applying graphlets to any graph $G$ is to enumerate the induced subgraphs so we can describe the local topology. For example, Figure 1 in \cite{Prz 1} shows all the induced undirected graphlets up to five nodes, which we include in the Appendix as Figure \ref{fig:graphlet_names}.  Observe that within each of these graphs $G_{s}$, some nodes are colored the same; this represents that they belong to the same orbit. For instance, from Figure \ref{fig:graphlet_names} one can see that $G_{1}$ has two orbits. This follows from the fact that $G_{1}$ has only two automorphisms, namely,
\begin{center}
    $Auto(G)=\{\mbox{ the identity function, the permutation $\alpha$ of the end points}\}$.
\end{center}
Hence, if $a,c$ denote the two endpoints and $b$ the only center point, the orbits are $O_{1}$ = $\{a,c\}$ and $O_{2}$ = $\{b\}$. Here, the endpoints are colored with black and the center node is colored with white. Likewise, $G_{2}$ has only one orbit because there is always an automorphism which sends one vertex to any other. Thus, the vertices are all in the same orbit $O_{3}$ signifying they all have equivalent topological positions.

\begin{figure}[h]  
\centering 
 \includegraphics[width=4.5in]{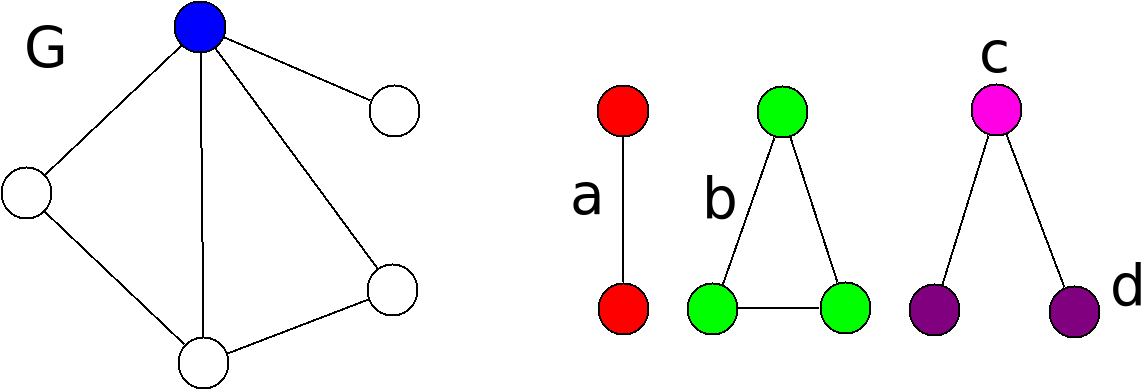}
 \caption{ The 4 different orbits within the graphlets $G_{0}, G_{1}, G_{2}$. Notice that each orbit is colored by a different color.}
 \label{fig:GDV}
\end{figure}

In \cite{Prz 2}, Pr\v{z}ulj introduces a graphlet-based similarity metric for undirected graphs called \textit{Graphlet Degree Distribution Agreement} (GDA) which was first tested in \cite{Prz 1} to study protein-protein Interaction networks. In that work the authors showed that the local network structure was related to specific biological functions. The GDA is constructed as follows: given a node $v$ of a network $G$, we account for the collection of all different graphlets $G_{s}$ in which vertex $v$ belongs to. For each step, we indicate the orbit or topological position within the graphlet $G_{s}$ the vertex $v$ is at. Then, we can construct a vector called the \textit{Graphlet Degree Vector} (GDV) which has dimension equal to the sum of all the orbits considered, i.e, the number of orbits in graphlet $G_{0}$ plus the number of orbits in graphlet $G_{1}$, etc. The entries of the vector reflect the number of times the chosen vertex $v$ belongs to the $i^{th}$ orbit within $G_{s}$. The graphlet degree vector of the vertex $v$, $GDV(v)$, contains local topological information about this vertex. As an example, consider the graph $G$ in Figure \ref{fig:GDV}. The vertex $v$ that is colored blue and the different orbits contained in the graphlets $G_{0}, G_{1}, G_{2}$ are drawn.

Thus, the graphlet degree vector associated with the blue vertex is:
\begin{equation*}
    GDV(v)=
    \begin{pmatrix}
    4\\
    2\\
    3\\
    0\\
    \end{pmatrix}
\end{equation*}

Repeating the process for for every vertex in the graph, one can construct a matrix whose rows are all the graphlet degree vectors. Furthermore, this matrix can be transformed into another matrix by computing the $s^{th}$-graphlet degree distribution followed by the arithmetic mean of all agreements over all the orbits considered. This matrix is called  the \textit{graphlet degree distribution}. It gives the number of nodes participating in graphlet $G_{s}$ $k$ times, and  has the property of storing not only the frequencies, but the distribution of the graphlets. 

\subsection{Directed networks}
For undirected graphs, the tuples $(u, v)$ and $(v, u)$ represent the same edge, meaning that the order of the pair $(u,v)$ is irrelevant. But when the order matters, in the sense that there is a directionality to the transfer of information between the vertex pair, then $(u, v)$ and $(v, u)$ represent different edges, each one of these representing a different direction. Graphs with this edge property are called \textit{directed graphs} or \textit{digraphs}. Directed edges represent a way of capturing asymmetric information flow in a network. For example, one can model metabolic reactions in Eukaryotes as an enzyme-enzyme network in which two enzymes-coding genes are connected by a directed line if the first enzyme catalyses a reaction whose product is a substrate for a reaction catalysed by the second enzyme; a biological example where graphlets have been used (see \cite{Prz 3}). Digraphs are an appropriate model for biological neural networks. For example, vertices can  represent neurons and directed edges represent axons and the directional propagation of action potentials between two neurons, or vertices can be brain regions connected by edges that represent white matter tracts, or even as a model of a neuron itself from the soma to axonal arborizations with vertices representing positions along the axon and edges representing axonal segments \cite{Puppo}.

\begin{figure}[h!]
    \centering 
    \includegraphics [width=3.5in]{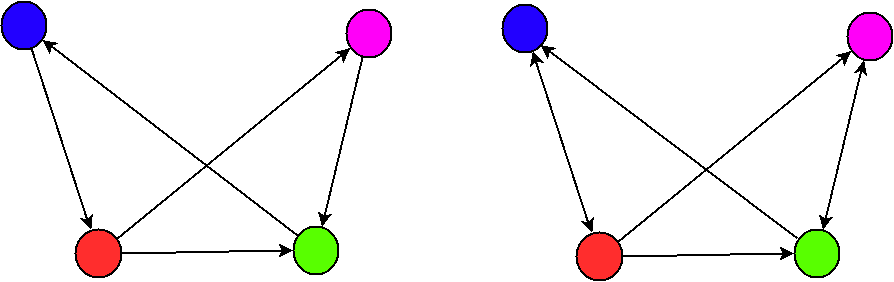}
    \caption{Directed and bidirected graphs.}
    \label{Figure 1.01}
\end{figure}

Two digraphs $G$ and $G'$ are isomorphic if there exists an isomorphism $f$ on their underlying graphs which preserves the direction of the edges. The automorphism group of a digraph $G$ is similar to an undirected graph, except now the group is smaller due to the directionality restrictions. This means the orbit sizes, the number of vertices with identical topology, is smaller. Inversely, the number of orbits is greater since there are more different types of vertex topological positions.  Nevertheless, the notion of graphlets and metrics in the directed case are still attainable. Given the set of all orbits for directed graphlets of size up to four nodes,  Pr\v{z}ulj, et. al. extended the Graphlet Degree Distribution Agreement (GDA) for directed graphs, showing its efficiency and superiority over existing network measures (see  \cite{Prz 3}).

\subsection{Temporal Networks}

Networks can also used to model temporal dynamics. These networks are interchangeably referred to as \textit{temporal networks} or \textit{dynamical networks}. A temporal network comprises a set of nodes $V$ and a set $E$ of events (temporal edges) that are associated with a start time and duration (see \cite{Milen}). Temporal networks who's dynamic are dependent on their spatial geometry in addition to their connectivity form a class of spatial-temporal networks that we have theoretically studied as models of biological neural networks \cite{Silva 2,Silva}. One way to study global dynamics is to aggregate all the nodes and edges from the temporal information into a single static graph. Another strategy is to represent the evolving dynamics by a sequence of consecutive network \textit{snapshots}, each of which is a static graph gathering all the temporal data observed during a time interval $T_{W}$. Static network metrics, such as GDA, have been used with temporal networks to measure how the topology changes over time by comparing consecutive snapshots. This is a static-temporal approach. Nevertheless, a static-temporal approach, and hence GDA, can toss out temporal information which is central for understanding the dynamic's evolution. In both cases though, the major shortcoming of these approaches is that they lose information about how the dynamics evolved over time

In \cite{Temp Net}, Aparicio, Ribiero, and Silva introduced a new metric for temporal networks called the  \textit{orbit-transition agreement} (OTA) which considers how local topologies change with time. They showed that OTA was able to show improved clustering accuracies of sets of temporal networks when compared to static network motifs and static graphlets.

\section{Synaptic Signal Graphs} \label{sec:signalgraphs}
The novel contribution of this paper is an extension of the theory of graphlets in order to allow us to analyze the causal evolution of dynamic patterns formed by our competitive refractory dynamics model. In the Introduction we briefly gave a high level overview of the framework. We refer the reader again to \cite{Silva} for details on the full development of the model and its theoretical analyses, and \cite{Silva 2} for background and results on the effects of network geometry on neural dynamics. In this section we extend this prior work in order to accommodate the graphlet analysis.

\subsection{Signal States}
\begin{figure}[htp]
    \centering
    \includegraphics[width=10cm]{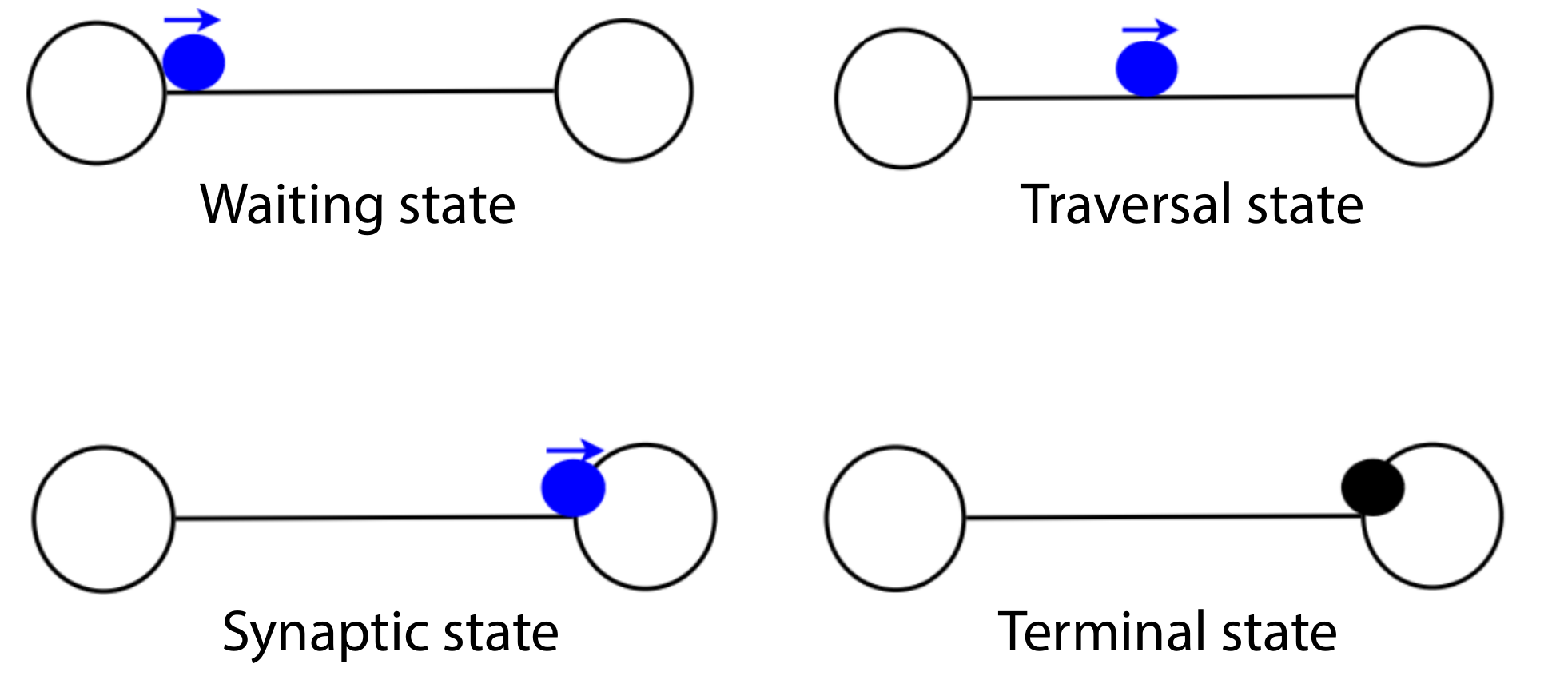}
    \caption{Possible dicrete signal states along an edge connecting two vertices (nodes) in a network. During the 'Waiting state' a signal has not yet been released by the initiating vertex. The 'Traversal state' reflects a signal in transit along the edge. The 'Active state' reflects the arrival of the signal at its downstream target vertex and a subsequent effect. In the 'Terminal state' the signal has had a past effect.}
    \label{fig:Signal States}
\end{figure}

Consider a discrete signal $\sigma$ sent from some start node $s(\sigma)$. This signal traverses down its geometric edge until it reaches its target node $t(\sigma)$. Each signal's "lifetime" can be described by a function $D(\rho,\psi,\alpha)$. These three parameters will facilitate deciphering the signal's causal effect within each of its states, namely if any effect exists. Figure \ref{fig:Signal States} shows the four possible states of a signal.

We consider the first state as a waiting period or \textit{waiting state} where the signal has not yet been sent. The signal necessarily has no effect on the dynamics. The second state begins when the signal's start node activates and fires, thus sending the signal towards its target node. We denote the time when $\sigma$ is sent from $s(\sigma)$ by $\rho(\sigma)$. This state is characterized by having $\sigma$ coming down the edge. We call this $\sigma$'s \textit{traversal state}. Observe that the traversal state is bounded by the time when $\sigma$ reaches its target node, which we denote by $\psi(\sigma)$. Also, during the traversal state, $\sigma$ does not cause any node to activate nor contributes any weight to the target's running summation. In other words, traversing has no \textit{effect} on any nodes, and therefore neither on the dynamics. It is only when the signal reaches its target node at time $\psi(\sigma)$ does it have an effect and contribute a weight to the running summation. If a signal reaches its target node but the node is refractory, then this signal is dropped, which is why its important to also mark $\sigma$ as entering its \textit{synaptic state} - analogous to the arrival of an action potential at a synaptic terminal in a biological neuron.  For the rest of the paper, we assume that we are only dealing with signals which do enter their synaptic state, since dropped signals do not have causal effects.

When a target node's running summation surpasses its activation threshold, at time $\alpha(\sigma)$, then the signal weight of $\sigma$ becomes irrelevant. This is because the target node becomes refractory and resets its running summation to zero, as required by the model. At time $\alpha(\sigma)$, the signal enters its \textit{terminal state}. Below we define the possible states of a signal given an observation time $T_{O}$ in the continuous time interval $[a,b]$.
\[
  \mathcal{S}_{T_{O}}(\sigma) =
  \begin{cases}
            waiting & \text{if $T_{O}$ $<$ $\rho(\sigma)$} \\
           
            traversal & \text{if $\rho(\sigma)$ $\leq$ $T_{O}$ $<$ $\psi(\sigma)$} \\
                    
            synaptic & \text{if $\psi(\sigma)$ $\leq$ $T_{O}$ $<$ $\alpha(\sigma)$}\\
            
            terminal & \text{if $\alpha(\sigma)$ $\leq$ $T_{O}$}

  \end{cases}
\]

\subsection{Signal Sets}

We want to investigate the interactions between signals and their collective effect for activating network nodes. We define the \textit{signal set} 
\begin{center}
    $\Phi$[$t_{a},t_{b}$] = $\{\sigma|t_{a} \leq \rho(\sigma) \leq t_{b}\}$. 
\end{center}
$\Phi$[$t_{a},t_{b}$] is the set of all signals sent from any network node within the time interval [$t_{a},t_{b}$] $\subseteq$ $\mathbb{R}$. By definition, it is clear that if [$t_{a},t_{b}$] $\subseteq$ [$t_{a},t_{c}$], then $\Phi$[$t_{a},t_{b}$] $\subseteq$ $\Phi$[$t_{a},t_{c}$]. Because there can only be a monotonic increase in the set of signals sent within the network as time passes, we can write the relation 
\begin{equation}
    \Phi[t_{a},t_{c}] = \Phi[t_{a},t_{b}] \sqcup \Phi(t_{b},t_{c}]
\end{equation}
In the event that $\Phi$[$t_{b},t_{c}$] is an empty set, the signal set remains constant from $\Phi$[$t_{a},t_{b}$] to $\Phi$($t_{b},t_{c}$]. For our purposes, we notice that the signal set undergoes a change or a transition when a node activates, because the next event involves this node sending signals to its downstream nodes. Also, we assume that there are finitely many time points before a node activates, hence finitely many transitions in the signal set. We can therefore enumerate these node activation times in increasing order as $\{t_{0},t_{1},\dots,t_{N-1},t_{N}\}$. Let $A$ be an order preserving map $A:\mathbb{N}\longrightarrow \mathbb{R}$ such that $A(j) = t_{j}$ for $j\leq N$ and $A(j) = t_{N}$ for $j>N$. We call the sequence $\{A(j)\}_{j \in \mathbb{N}}$: = $\{A_{j}\}_{j \in \mathbb{N}}$ the \textit{activation series} and a particular $A_{j}$ an \textit{activation-step}. As a result, we can construct the sequence of signal sets $\Phi_{\subseteq}$: 
\begin{equation}\label{Signal Set Sequence}
\Phi[0,A_{0}] \subseteq \Phi[0,A_{1}] \subseteq \Phi[0,A_{2}]\subseteq \dots \subseteq \Phi[0,A_{N-1}] \subseteq \Phi[0,A_{N}] \subseteq \dots
\end{equation}
This construction preserves all the pertinent causal information in the dynamics, while simultaneously supporting what will be the requirements of a graphlet-based temporal analysis. Recall that the definition of a snapshot-based representation introduced in \cite{Milen} and \cite{Temp Net} requires snapshots to be equally spaced apart by $\Delta t$. However, here the set of all activation-steps do not necessarily need to be separated by a fixed time interval. Thus, we generalize our window sizes to allow them to change rather than keeping them fixed. Specifically, we chose to take such generalized snapshot-based representations of signal dynamics at each activation-step.

Before continuing, we make an important observation. Suppose we had two identical sequences of signal sets $\Phi_{\subseteq}^{1}$ and $\Phi_{\subseteq}^{2}$. It is not necessary that the actual signal dynamics are the same. To demonstrate this point, take two signal dynamics $\mathbb{D}_{1}$ and $\mathbb{D}_{2}$, along with their corresponding sequences such that $\Phi_{\subseteq}^{1}(\mathbb{D}_{1}) = \Phi_{\subseteq}^{2}(\mathbb{D}_{2})$. We can think of these two signal dynamics as films that can be played at three different speeds: slow, normal, or fast. Whether we watch the same film at slow, normal or fast, we still see the same succession of scenes or actions. Thus, the similarity or invariance between the films is the sequence of scenes and its difference is their respective \textit{time lapse}. Hence, if two signal dynamics $\mathbb{D}_{1}$ and $\mathbb{D}_{2}$ have the same sequence of signal sets, we say the signal dynamics are equivalent up to time lapse. This is a byproduct of interrogating causal relationships embedded in possibly variable dynamics.

\subsection{Synaptic Signal Sets and Graphs}

In this section we describe a snapshot-based representation of the competitive refractory dynamic model. We do this by constructing synaptic signal graphs $G_{\Psi}$ to play the role of snapshots. In this way, we will be able to represent signal dynamics as a sequence of synaptic signal graphs. 

Given a time interval $[t_c,t_d] \subseteq [t_a,t_b]$, we define the \textit{synaptic signal set} 
\begin{center}
    $\Psi[t_{c},t_{d}]$ = $\{\sigma \in \Phi[t_{a},t_{b}] | t_{c} \leq \psi(\sigma) \leq t_{d} \leq \alpha(\sigma)\}$. 
\end{center}
By construction, it follows that $\Psi[t_{c}, t_{d}]$ $\subseteq$ $\Phi[t_{a}, t_{b}]$. Using the activation series $\{A_{j}\}_{j \in \mathbb{N}}$ we can construct the sequence of synaptic signal sets:
\begin{center}
    $\Psi[0,A_{0}], \Psi[0,A_{1}], \Psi[0,A_{2}],\dots, \Psi[0,A_{N-1}],\Psi[0,A_{N}],\dots$
\end{center}
Let us take a look at a small subsequence: 
\begin{center}
    $\dots,\Psi[0,A_{j-1}], \Psi[0,A_{j}], \Psi[0,A_{j+1}],\dots $
\end{center}
The synaptic signal set $\Psi[0,A_{j}]$ can be partitioned in two ways, both resulting in two disjoint sets. The first partition classifies signals via their relation to the node(s) activated at time $A_{j}$. Let the set $\Psi_{act}[0,A_{j}]$ consist of all the signals whose target node activates at $A_{j}$, while the second set $\Psi_{act}^{c}[0,A_{j}]$ has all the other synaptic signals whose target node does not activate. Thus, we have the disjoint union
\begin{equation}\label{act part}
    \Psi[0,A_{j}] = \Psi_{act}[0,A_{j}] \  \dot\sqcup \  \Psi_{act}^{c}[0,A_{j}].
\end{equation}

A second partition of $\Psi[0,A_{j}]$ is due to the signals in $\Psi_{act}^{c}[0,A_{j-1}]$ which remain synaptic in the next activation-step, whereby they are a subset in $\Psi[0,A_{j}]$. Hence,
\begin{equation}\label{new part}
    \Psi[0,A_{j}] = \Psi_{act}^{c}[0,A_{j-1}] \  \dot\sqcup \  \Psi_{new}(A_{j-1},A_{j}],
\end{equation}
where $\Psi_{new}(A_{j-1},A_{j}]$ is the set of new synaptic signals. That is, 
\begin{center}
    $\Psi_{new}(A_{j-1},A_{j}] = \{\sigma| A_{j-1} < \psi(\sigma) \leq A_{j} \leq \alpha(\sigma)\}$.
\end{center} 
As a consequence, for the subsequence we have:
\begin{equation} \label{Embedding}
\begin{split}
\Psi[0,A_{j+1}] & = \Psi_{act}^{c}[0,A_{j}] \  \dot\sqcup \  \Psi_{new}(A_{j},A_{j+1}] \\
 & = \Big(\Psi[0,A_{j}] \backslash \Psi_{act}[0,A_{j}]\Big) \  \dot\sqcup \  \Psi_{new}(A_{j},A_{j+1}] \\ 
 & = \Big(\big(\Psi_{act}^{c}[0,A_{j-1}] \  \dot\sqcup \ \Psi_{new}(A_{j-1},A_{j}]\big) \backslash \Psi_{act}[0,A_{j}]\Big) \  \dot\sqcup \  \Psi_{new}(A_{j},A_{j+1}]
\end{split}
\end{equation}
Intuitively, this means that there is subset from $\Psi_{act}^{c}[0,A_{j-1}]$ which remains in $\Psi[0,A_{j+1}]$. 

From any synaptic signal set $\Psi[0,A_{j}]$, we can create a snapshot which is a graph that encodes the respective set of synaptic signals. We will refer to these graphs as \textit{synaptic signal graphs} and write them as $G_{\Psi}[A_{j}]$. The edge set of any $G_{\Psi}[A_{j}]$ corresponds to the synaptic signals in $\Psi[0,A_{j}]$ and the vertex set is comprised of all the network nodes.

The above synaptic signal set decompositions have a correspondence to their synaptic signal graph counterparts. It is possible for there to be subgraphs of $G_{\Psi}[A_{j-1}]$ which are embedded into $G_{\Psi}[A_{j+2}]$. This is because synaptic signals may remain synaptic through multiple network node activations without their downstream nodes activating. To achieve this, we let $H_{\Psi}[A_{j}]$ correspond to the synaptic signal subgraph of the synaptic signal subset $\Psi_{act}^{c}[0,A_{j}]$ $\subset$ $\Psi[0,A_{j}]$. 

In the next section we how how synaptic signal graphs are represented by \textit{multidigraphs}, which allow us to accurately express a full description of the dynamics by writing down their edge-ordering relating to the order in which the signals became synaptic. A graphlet-based analysis then follows by constructing the appropriate graphlet types.

\subsection{Synaptic Signal Multidigraphs}

In some cases,  a node $u$ contributing to the running summation of a node $v$ could could consist of multiple synaptic signals contributing towards a single activation event. This happens when the upstream node $u$ activates multiple times between two consecutive activations of node  $v$ such that at least two signals of $u$ become synaptic before the second activation of $v$. In order to preserve the  property of synaptic signals multiplicity, we use multidigraphs. Thus, we will write each synaptic signal graph as a multidigraph. 

To facilitate notation, for the remainder of the w papere will denote a synaptic signal multidigraph at $A_{j}$ as $M_{\Psi}[A_{j}]$. Again, for each $M_{\Psi}[A_{j}]$ the vertex set $V(M_{\Psi}[A_{j}])$ equates to the network nodes whereas the edge set corresponds to the set of all synaptic signals in $\Psi [0,A_{j}]$. In this context, there will be directed edges $(u,v)$ of degree $d$ in $E(M_{\Psi}[A_{j}])$ if and only if there exists $d$ synaptic signals $\sigma_{i}$ from $u$ to $v$ with $0 \leq i \leq d$ and $A_{j} < \psi(\sigma_{i}) \leq A_{j+1}$ $<$ $\alpha(\sigma_{i})$ for $i \in \{1,2,\dots, d\}$. We will write the degree of directed edges as $d_{u,v}$ where $u$ denotes the start vertex and $v$ the target vertex of these edges. 

Given a graph representation by $M_{\Psi}[A_{j}]$ for each snap shot, we can now analyze it by a graphlet decomposition. However, existing graphlets do not capture information about multiple directed edges. Rather than make an enumeration of the graphlets (for each of any occurring multiple edges), we will take a different approach. We will construct a vector space as a means to describe and extend directed graphlets to directed multigraphlets in the same framework. We first start by viewing directed graphlets as linear combinations of directed edges, and then generate directed multi-graphlets or multidigraphlets by assigning coefficients to these linear combinations whose values are equal to the degrees of the multiple directed edges. The advantage of having this algebraic construction of multidigraphlets from directed graphlets is two-fold. First, it is a natural mapping from the automorphism orbits of directed graphlets to those of multidigraphlets. This not only gives a relationship between the orbits, but also topological relationships. The second advantage is it provides a fine-tuned enumerating process of multidigraphlets.  

\subsubsection{Algebraic Representation of Multidigraphlets}

We begin by describing how to algebraically represent and construct directed graphlets via their edges. This construction is based on the concept of \textit{elementary paths} as in \cite{Quiv}. We adapt the following definition to our framework for a multidigraph $M$: 
\begin{defi} \cite[Definition 2.2]{Quiv}    
Given a positive integer $r$, an elementary $r$-path in a multidigraph $M$ is a non-empty sequence $a_{0},a_{1}, \cdots a_{r-1}$ of edges in $M$ such that $t(a_{i})=s(a_{i+1})$ for $i\in \{1,\cdots,r-2\}$. We denote this $r$-path by $p=a_{0}a_{1} \cdots a_{r-1}$, whose start vertex $s(p)=a_{0}$ and target vertex $t(p)=a_{r-1}$.
\end{defi}
Notice that for $r=0$, the $0$-elementary paths have the form $p=v$, where $v\in V(M)$ is some vertex of the multidigraph $M$ and when $r=1$, the $1$-elementary paths are comprised by all the directed edges in $E(M)$. Let $K$ be an integral domain and let us denote by $\Lambda_{0}$ the set of all linear combinations of $0$-elementary paths and by $\Lambda_{1}$ the set of all linear combinations of $1$-elementary paths in $M$ over $K$ respectively. Graphically, an element in $\Lambda_{0}$ translates to a finite set of isolated vertices, whereas directed and multidirected subgraphs in $\Lambda_{1}$ are a linear combination of those vertices. Thus we define $\Lambda=\Lambda_{0}\oplus \Lambda_{1}$, whose elements are all linear combinations generated by both $0$-elementary paths and $1$-elementary paths. As we will see, all possible states in which a multidigraph can be found can be represented by a linear combination in $\Lambda$.

According to the directed graphlets enumeration in \cite{Extending}, graphlet $G_{0}$ corresponds to a single directed edge from node $v_{0}$ to node $v_{1}$. Thus, its algebraic representation is $1(v_{0},v_{1})$. So, if we start with the vertex set $V = \{v_{0},v_{1}\}$, the graphs $G^{0}, G^{1}$ with edge sets $E^{0} = \{(v_{0},v_{1})\}$ and $E^{1} = \{(v_{1}v_{0})\}$, respectively, both belong to the graphlet type $G_{0}$. In the first case $v_{0}$ belongs to orbit $O_{0}$, and $v_{1}$ belongs to orbit $O_{1}$. The second case is the reverse. For graphlet type $G_{1}$ (a bidirectional edge) the representation is $1(v_{0},v_{1}) + 1(v_{1},v_{0})$, where $v_{0},v_{1}$ belong to the same orbit $O_{2}$. Suppose now that we have a graph $G^{2}$ whose edge set consists of $d$ edges $(v_{0},v_{1})$. We represent this multiplicity with the coefficient $d_{v_{0},v_{1}} \in \mathbb{N}$, such that $G^{2}$ is expressed by $d_{v_{0},v_{1}}(v_{0},v_{1})$. The orbits of this graph are closely linked to the orbits of $G_{0}$, but could be different. Likewise, if we have a graph $G^{3}$ whose edge set had $d$ edges $(v_{0},v_{1})$ and $d$ edges $(v_{1},v_{0})$, the orbits will have the same vertices as elements. However, if $d_{v_{0}v_{1}} \neq d_{v_{1}v_{0}}$, then the vertices will not be in the same orbit because the symmetry breaks as a result of the different degrees of the directed multi-edges. Thus, the graphlet type is expressed as $d_{v_{0}v_{1}}(v_{0},v_{1}) + d_{v_{1}v_{0}}(v_{1},v_{0})$ with $v_{0}$ and $v_{1}$ belonging to distinct orbits. Note that the expression is unique up to reordering. These algebraic expressions can be combined to express any multidigraph, and in particular any multidigraphlet type. This example demonstrates the ability of graphlets to be combined together to produce new graphlets. 

\begin{figure}[h]
    \centering
    \includegraphics[width=3.5in]{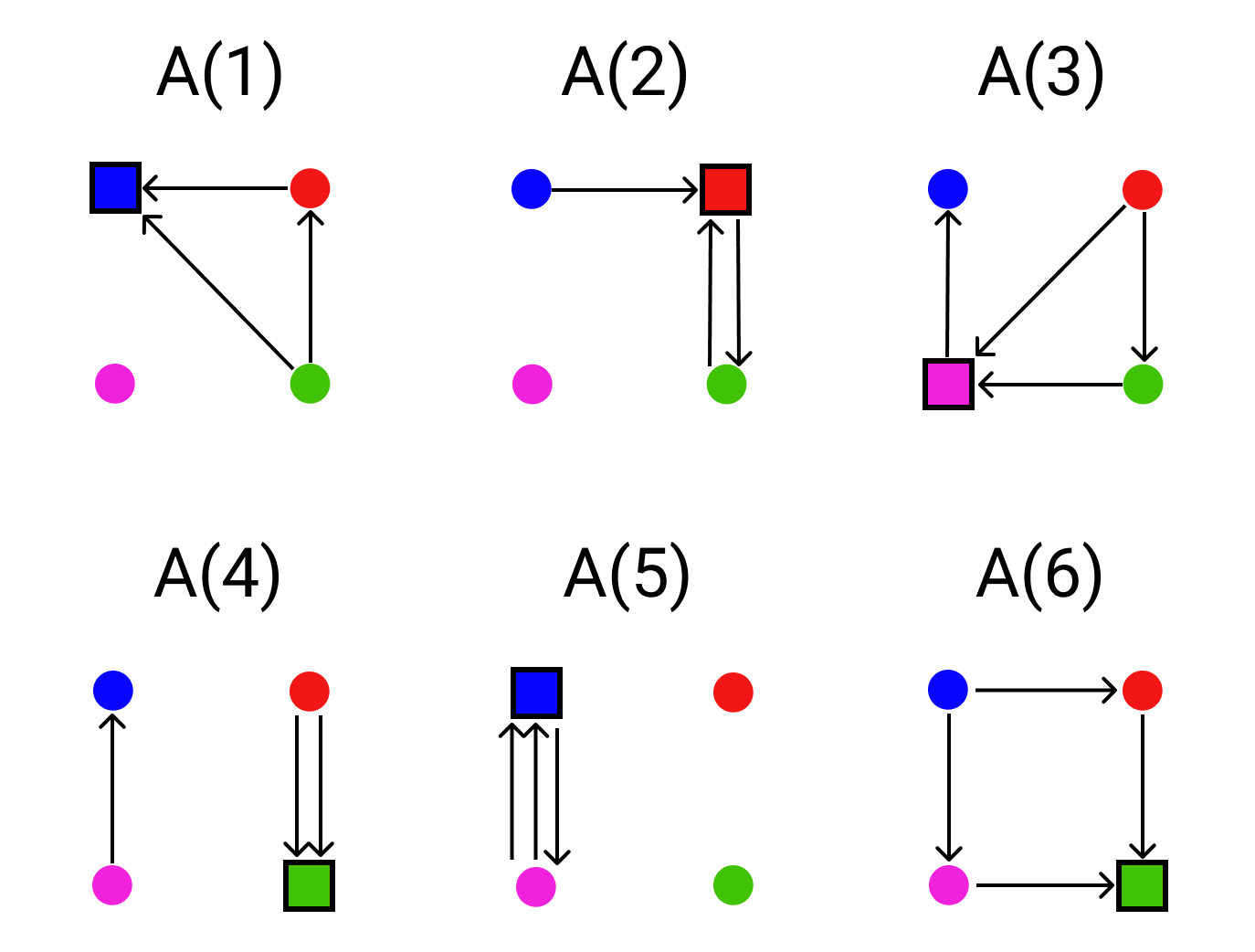}
    \caption{Resulting multidigraphs after applying the node equivalence relation between vertices of a DAG.}
    \label{fig:MDGs}
\end{figure} 

\begin{defi}
We say that two multidigraphs $M_{1}$ and $M_{2}$ are isomorphic if there exists a bijection $f:V(M_{1})\longrightarrow V(M_{2})$ and a bijection $g:E(M_{1})\longrightarrow E(M_{2})$ such that for every directed multi-edge $(u,v)$ $\in$ $E(M_{1})$, we have that $g(u, v) = (f(u), f(v)) \in M_{2}$ and $d_{u,v} = d_{f(u)f(v)}$. 
\end{defi}
In other words, any multidigraph isomorphism is an isomorphisms between the underlying directed graphs which preserves the edge multiplicities. In a follow up paper we will introduce the use of this construction as part of an algorithm that will allow us to build up dynamic patterns produced by the competitive refractory  model.

\subsubsection{Examples of multidigraphlets}

In the same way that the automorphism orbits for directed graphs decreases their number of elements, the vertex orbits of multidigraphs have fewer elements in each orbit equivalence class, but more classes in total. This is because most of the automorphism groups are trivial; that is, they only contain the identity map. Consequently, we lose a lot of the symmetry associated with undirected and directed graphlets. Nevertheless, despite the loss of symmetries, orbits can still be used to indicate which topological position a vertex or edge belongs to within a multidigraphlet.

Consider the snapshots taken at $A(4)$ and $A(5)$ in Figure $\ref{fig:MDGs}$. Since the other snapshots are merely directed graphs, their graphlets are also directed. For this reason, we work through $M_{\Psi}(A_{4})$ and $M_{\Psi}(A_{5})$ to explain the new case of multidigraphlets. 

\begin{figure}[h] 
\begin{subfigure}{0.5\textwidth}
    \includegraphics[width=1.0\linewidth]{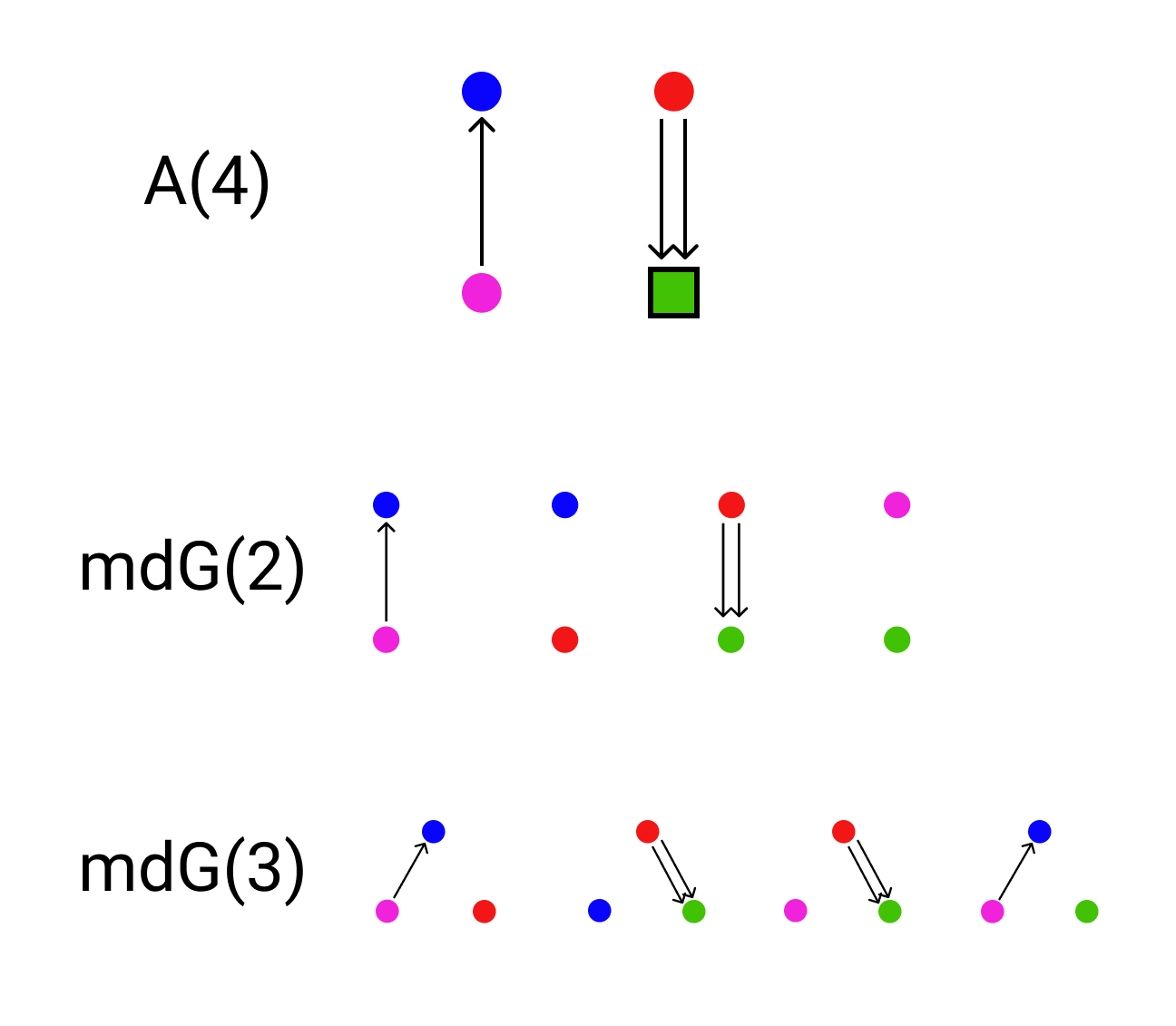}
    \caption{Graphlets of $M_{\Psi}(A_{4})$.}
    \label{fig:Graphlets1}
\end{subfigure} 
\begin{subfigure}{0.5\textwidth}
    \includegraphics[width=1.0\linewidth]{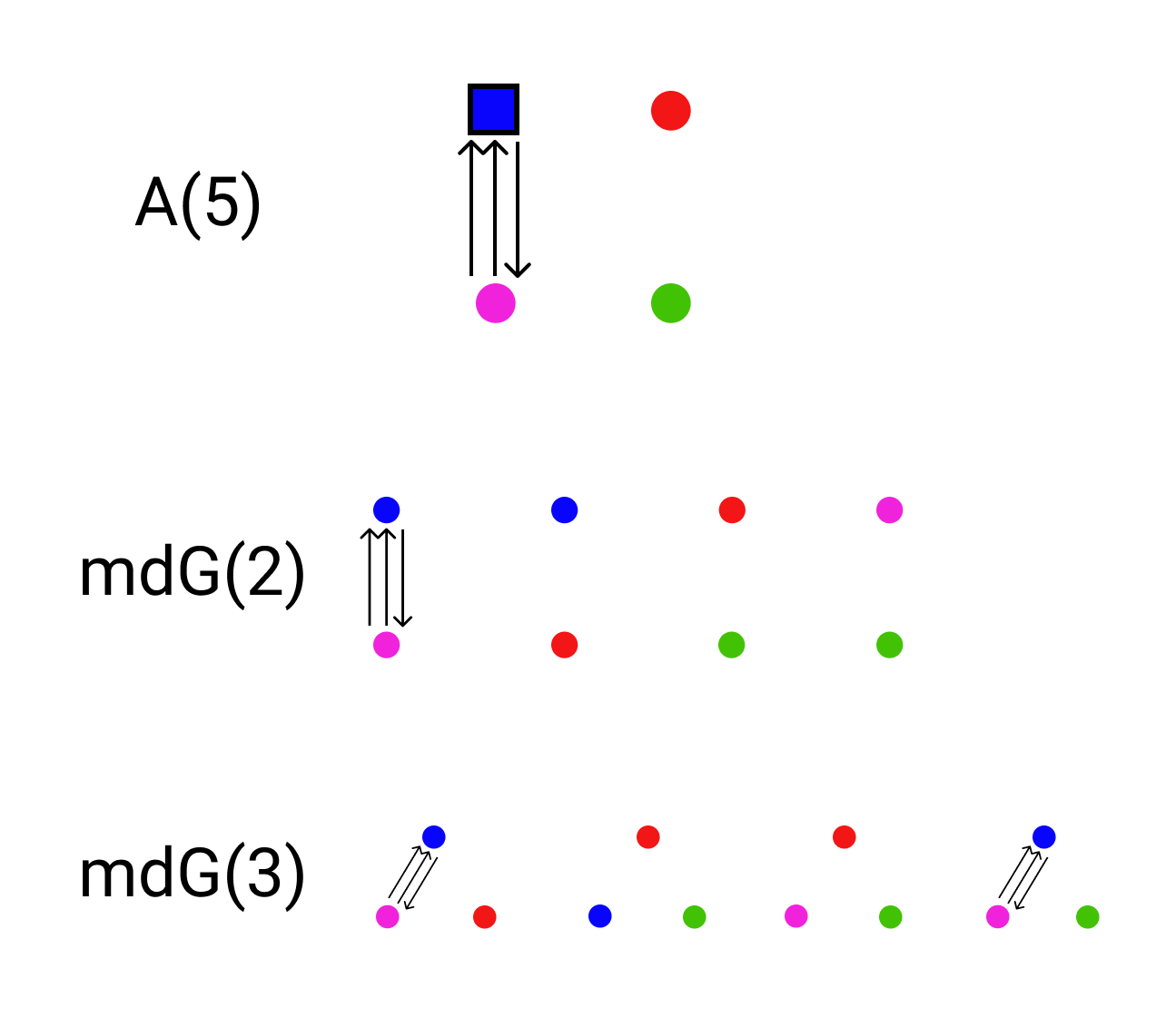}
    \caption{Graphlets of $M_{\Psi}(A_{5})$.}
    \label{fig:Graphlets2}
\end{subfigure}
\caption{The two node and three node decomposition of $M_{\Psi}(A_{4})$ and $M_{\Psi}(A_{5})$.}
\label{fig:Multidigraphlets Examples}
\end{figure}

Figure $\ref{fig:Multidigraphlets Examples}$ illustrates the two and three node graphlets of $M_{\Psi}(A_{4})$ and $M_{\Psi}(A_{5})$, respectively. The algebraic expressions for the two and three nodes graphlets of $M_{\Psi}(A_{4})$ can be written algebraically as 
\begin{center}
    $mdG_{2}$ = $\{(P,B); B + R; 2(R,G); P + G\}$\\
\vspace{2mm}    
$mdG_{3}$ = $\{R + (P,B); B + 2(R,G); 
P + 2(R,G); G + (P,B)\}$
\end{center}
And the graphlets of $M_{\Psi}(A_{5})$ are
\begin{center}
    $mdG_{2}$ = $\{2(P,B) + (B,P); B + R; R + G; P + G\}$\\
\vspace{2mm}   
$mdG_{3}$ = $\{R + [2(P,B) + (B,P)]; R + B + G; R + P + G; G + [2(P,B) + (B,P)]\}$.
\end{center}

\subsubsection{Multidigraphlets catalog}

We introduce a three parameter characterization to catalog submodule spanning multidigraphlets. Let $\mathcal{V}$ = $|V'(M_{\Psi})|$ be the number of vertices a multidigraphlet has, $\mathcal{E}$ = $|E'(M_{\Psi})|$ be the number of network edges selected, and $\mathcal{F}$ the number of synaptic signals between the network nodes incident to the selected edges. Given values for $\mathcal{V}$, $\mathcal{E}$, and $\mathcal{F}$, we can output a set of multidigraphlets $M(\mathcal{V}, \mathcal{E}, \mathcal{F})$. For example, suppose $\mathcal{V}=3$, $\mathcal{E}=4$, and $\mathcal{F}$ arbitrary.  The possible underlying directed graphlets (up to isomorphisms) connecting these three vertices are shown in Figure $\ref{M(3,4,F)}$. The algebraic expressions for these four multidigraphlets representatives then are given by 
\begin{enumerate}
    \item $d_{RB}(RB) + d_{BR}(BR) + d_{RG}(RG) + d_{GR}(GR)$
    \item $d_{BR}(BR) + d_{GR}(GR) + d_{BG}(BG) + d_{GB}(GB)$
    \item $d_{RB}(RB) + d_{RG}(RG) + d_{BG}(BG) + d_{GB}(GB)$
    \item $d_{RB}(RB) + d_{BR}(BR) + d_{RG}(RG) + d_{GB}(GB)$
\end{enumerate}  
However, it is important to note that the sum of the coefficients  $d_{i,j}$ of each expression is always equal to $\mathcal{F}$:

\begin{figure}[h]
    \centering
    \includegraphics[width=4in]{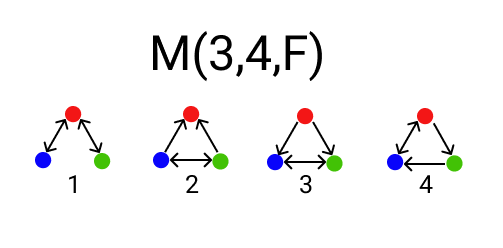}
   \caption{The four possible four underlying directed edges between three nodes which to span the class of multidigraphlets $M(3,4,\mathcal{F})$. The sum of all the edge coefficients must add up to $\mathcal{F}$.}
    \label{M(3,4,F)}
\end{figure}

\subsection{Edge-Ordered Synaptic Signal Multidigraphs}
The competitive refractory dynamic model treats weight contributions triggered by arriving signals from upstream nodes such that the maximum value of an edge weight to the running summation results at the time of arrival of the signal, with progressively decaying contributions at subsequent time steps (\emph{c.f.} equations \ref{eq:w1} and \ref{eq:decayeq} above). The physiological analog of this process is the decay of post-synaptic potentials as a function of the  space and time decay constants due to the membrane biophysics. This reflects a critical algorithmic computational component of the neurobiology.

\begin{figure}
    \centering
    \includegraphics[width=15cm]{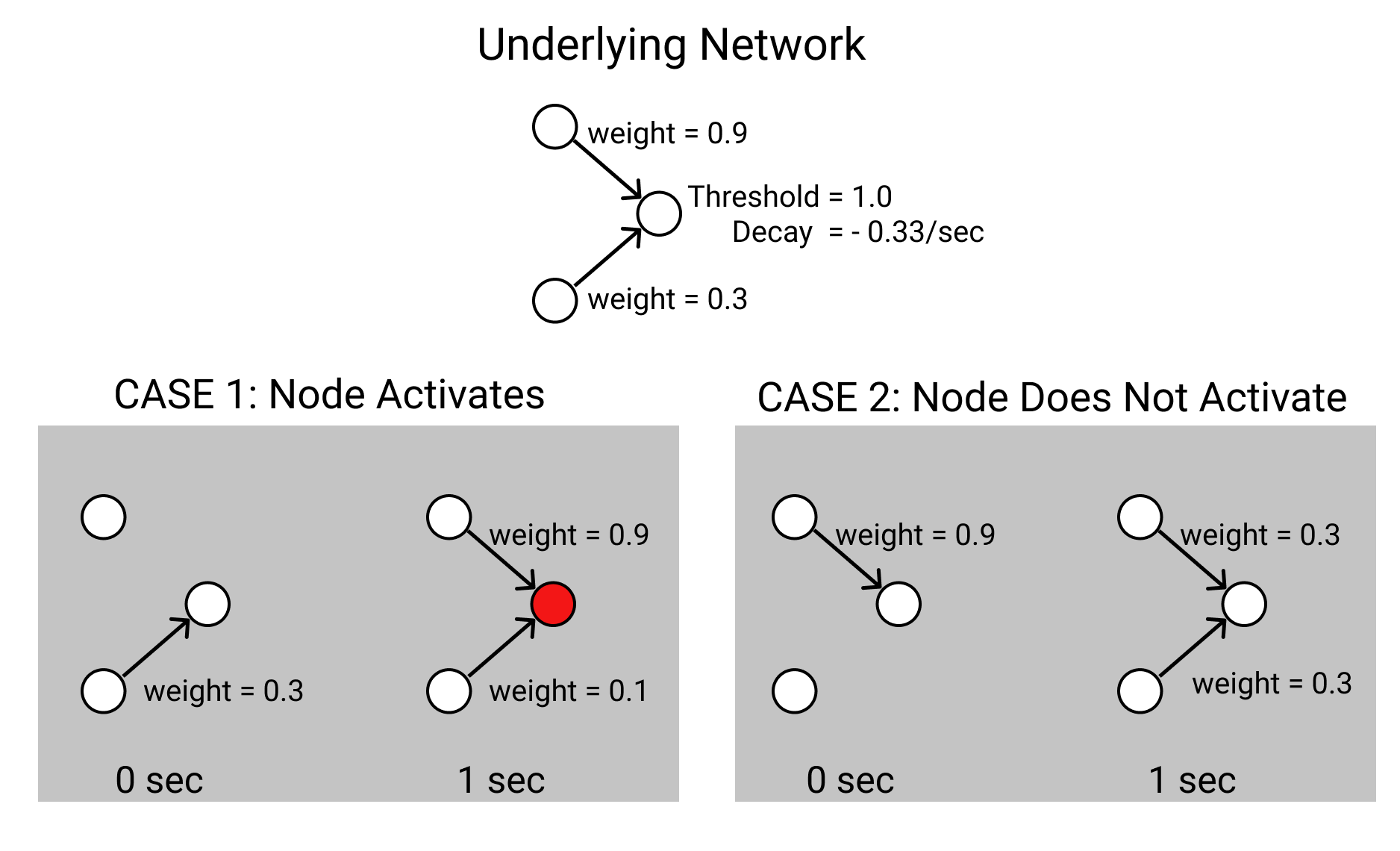}
    \caption{Two signals approach the target node. The signal weights are 0.9 and 0.3, respectively. For the target node, the activation threshold is 1.0 and the decay function is -0.33 of the signal weight each second. In case 1, the signal with weight equal to 0.3 decays first. A second later, the decay function brings the signal weight to 0.1. This is sufficient to be added with the second signal with weigth equal to 0.9 to overcome the activation threshold. In case 2, because the signal with weight 0.9 came first it decays to 0.3. The sum of the two signals is not sufficient to overcome the threshold, thus no activation results. }
    \label{fig:OrderMatters}
\end{figure}

Because of this, it is important to be keep track of the order in which the signals become synaptic. This motivates an \textit{edge-ordering} on the multidigraphs $M_{\Psi}$. As an example, figure $\ref{fig:OrderMatters}$ compares two cases with the same signals, but shows how the relative order affects how a node activates. 

The sequence of signal sets is our starting point. Referring back to sequence ($\ref{Signal Set Sequence}$), for each signal set $\Phi[0,A_{j}]$, we apply a total order relation $\leq$ \footnote{ A total order on a set $S$ is a binary relation $\leq$  such that the following properties hold: Reflexivity, Antysimmetry, Transitivity and Comparability.} based off the times they became synaptic, that is, for any $\sigma_{k}, \sigma_{l} \in \Phi[0,A_{j}]$, then $\sigma_{k} \leq \sigma_{l}$ if $\psi{(\sigma_{k})} \leq \psi(\sigma_{l})$.
We will denote each totally ordered signal set as $\Phi_{\leq}[0,A_{j}]$ for each $j\in \mathbb{N}$. Now, given $\Phi_{\leq}[0,A_{j}]$, we will denote by  $\Psi_{\leq}[0,A_{j}]$ the induced totally ordered synaptic signal set. This new condition allows us to re-describe the two sets as monotonic increasing sequences $P(\Phi_{\leq}[0, A_{j}])$ and $P(\Psi_{\leq}[0, A_{j}])$. In fact, for the sequence $P(\Phi_{\leq}[0, A_{j}])$ = $\sigma_{1}, \sigma_{2}, \dots,  \sigma_{N}$, the corresponding synaptic signal sequence $P(\Psi_{\leq}[0, A_{j}])$ is a subsequence $\sigma_{k_{1}}, \sigma_{k_{2}}, \dots, \sigma_{k_{S}}$, where $N \geq S$ are the number of signals in $\Phi_{\leq}[0, A_{j}]$, and $\Psi_{\leq}[0, A_{j}]$, respectively \cite{Rudin}.  An important observation is that the last synaptic signal in $\Psi_{\leq}[0, A_{j}]$ refers to the signal(s) which caused its target node(s) to activate. 

Let us consider now a multidigraph $M$. We can give two notions of an edge-ordering of $M$ from two perspectives. An \textit{absolute edge-ordering} is defined as a bijection $p_{abs}: E(M) \longrightarrow \{k_{1}, k_{2}, \dots, k_{|E(M)|}\}$ such that $k_{i} \leq k_{j}$ whenever $i < j$. A \textit{relative edge-ordering} is a bijection $p_{rel}: E(M) \longrightarrow \{1, 2,\dots, |E(M)|\}$ \cite{EOGraphs}. Observe that one can go from an absolute edge-ordering to a relative one, just by sending $k_{i}$ to $i$. Now, given a sequence of edges $P = e_{1}, e_{2},\dots, e_{l}$ we say that $P$ is \textit{monotonically increasing sequence of length l} if it is a sequence in $M$ such that $p(e_{i}) \leq p(e_{j})$ for all $i < j$. We denote an \textit{edge-ordered multidigraph} as an ordered pair $(M, p)$, where $M$ is a multidigraph and $p$ is an absolute or a relative edge-ordering.

To construct an \textit{edge-ordered synaptic signal multidigraph} $(M_{\Psi}[A_{j}], p)$, we first notice that for an absolute edge-ordering, $p_{abs}^{-1}(k_{i}) = (s(\sigma_{k_{i}}), t(\sigma_{k_{i}})) \in E(M_{\Psi}[A_{j}])$; that is, the edges get assigned the absolute ordering of the signal. In contrast, for a relative edge-ordering, $p_{rel}^{-1}(i) = (s(\sigma_{k_{i}}), t(\sigma_{k_{i}})) \in E(M_{\Psi}[A_{j}])$ -  the edge-ordering preserves the relative order in which the signals became synaptic. This is a crucial difference compared to dynamic graphlets. Dynamic graphlets have a relative edge-ordering which represents consecutive causal events. Dynamic graphlets are essentially causal paths where the sequence of edges must have vertices incident to consecutive edges. In our construction, we loosen this requirement. In fact, no causal relations exist within an edge-ordered multidigraph. This is because every time a signal activates its target node, the representative edge is removed in the next snapshot since the signal is no longer synaptic by definition. As a consequence, there exists no causal sequence within any ($M_{\Psi}[A_{j}], p$). The causal sequences are seen between different snapshots ($M_{\Psi}[A_{j}], p$) and ($M_{\Psi}[A_{j+c}], p$) where $c>0$. 

\begin{defi}
An \textit{edge-ordered multidigraph isomorphism} is a multidigraph isomorphism $f: V(M_{1}) \longrightarrow V(M_{2})$ such that for the edge-orderings $p_{abs}^{1}$ of $M_{1}$ and $p_{abs}^{2}$ of $M_{2}$, we have $p_{abs}^{1}(xy)$ = $p_{abs}^{2}(f(x)f(y))$.
\end{defi}
 
\begin{figure}[h!]
\begin{subfigure}{0.9\textwidth}
    \centering
    \includegraphics[width=5cm]{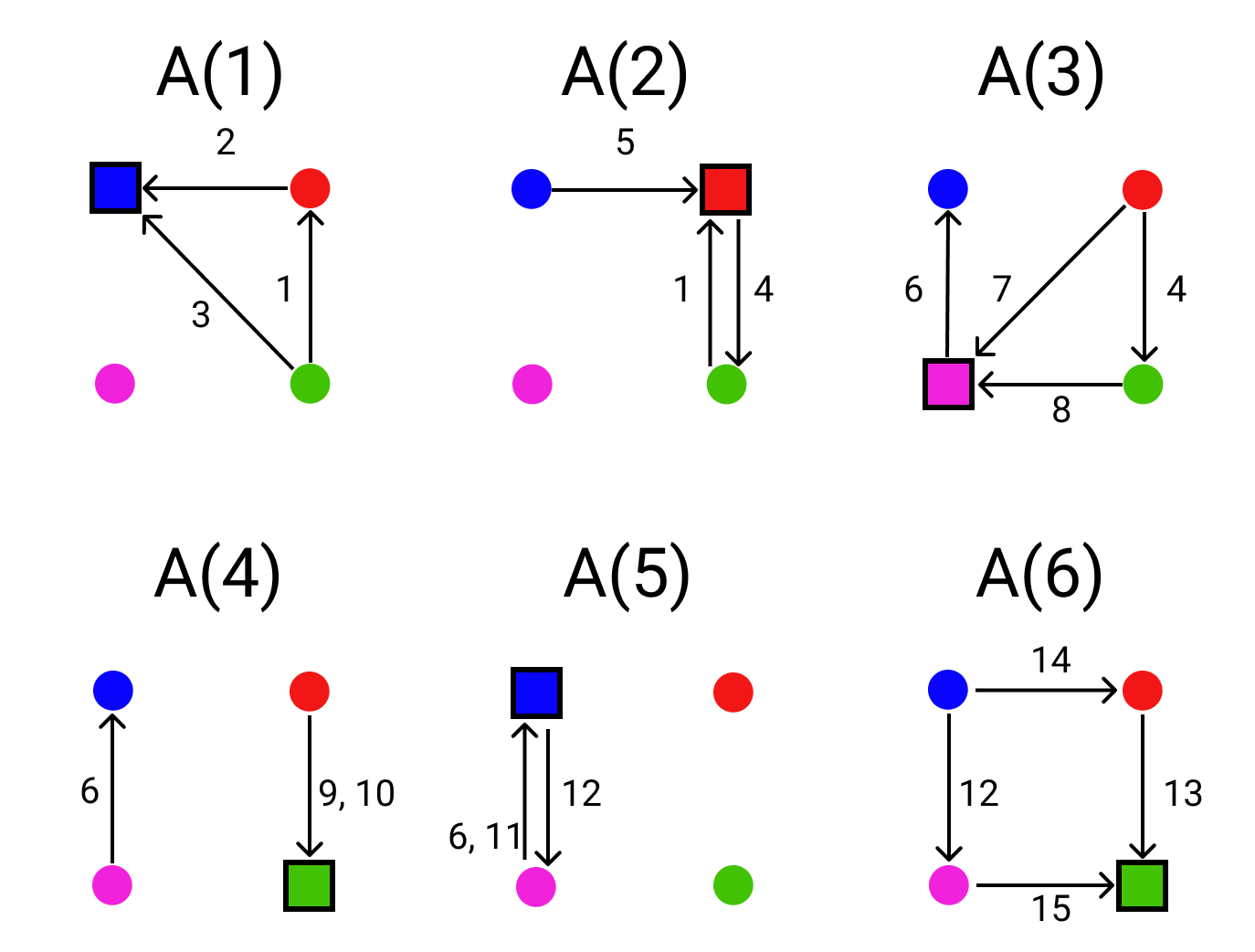}
    \caption{Edge-ordered multidigraphs. We suppress the multiple edges to conserve space.}
    \label{fig:EOMDG}
\end{subfigure}
\begin{subfigure}{0.45\textwidth}
    \includegraphics[width=1.0\linewidth]{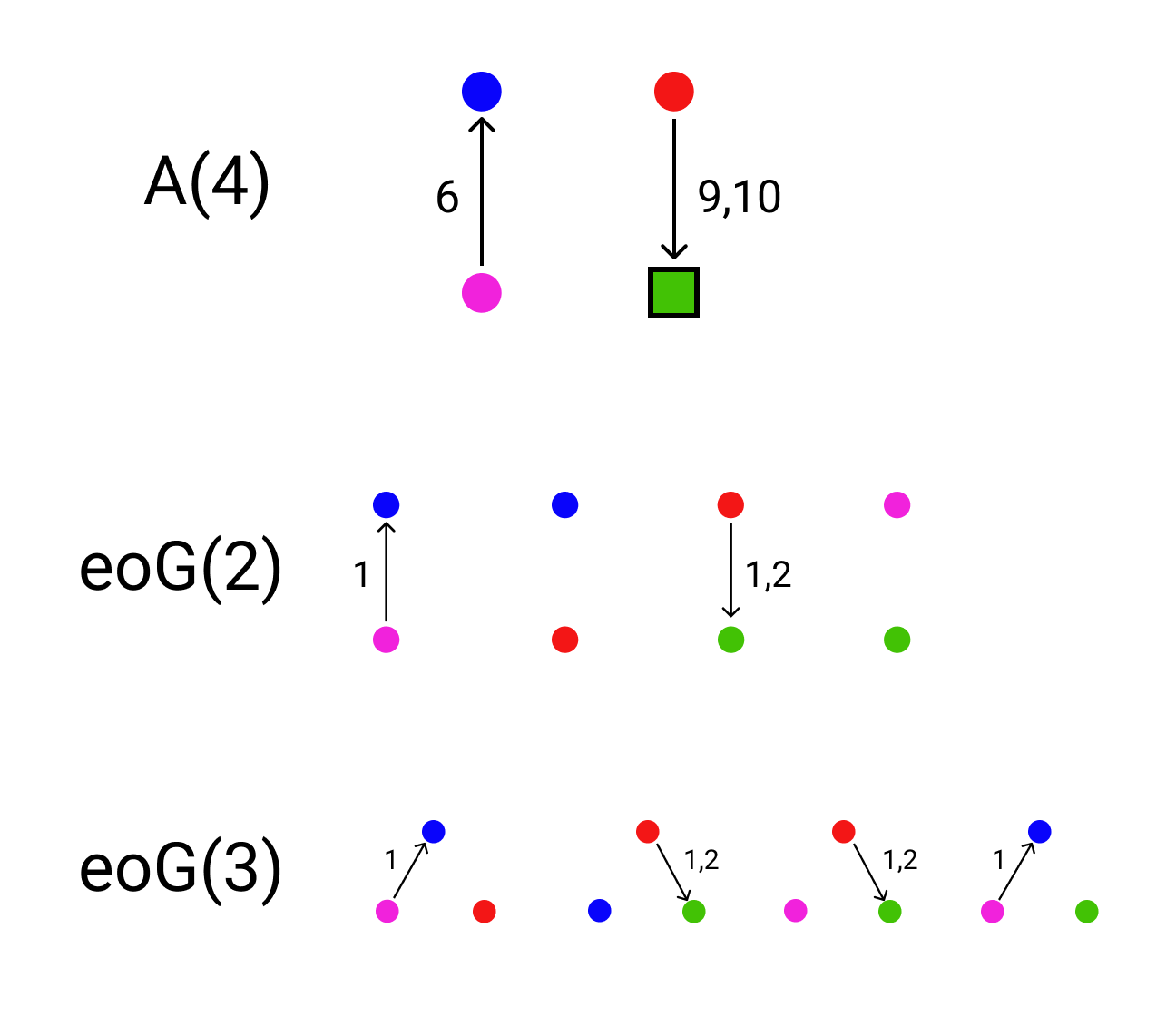}
    \caption{Graphlets of $(M_{\Psi}[A_{4}],p)$.}
    \label{fig:EOGraphlets1}
\end{subfigure} 
\begin{subfigure}{0.45\textwidth}
    \includegraphics[width=1.0\linewidth]{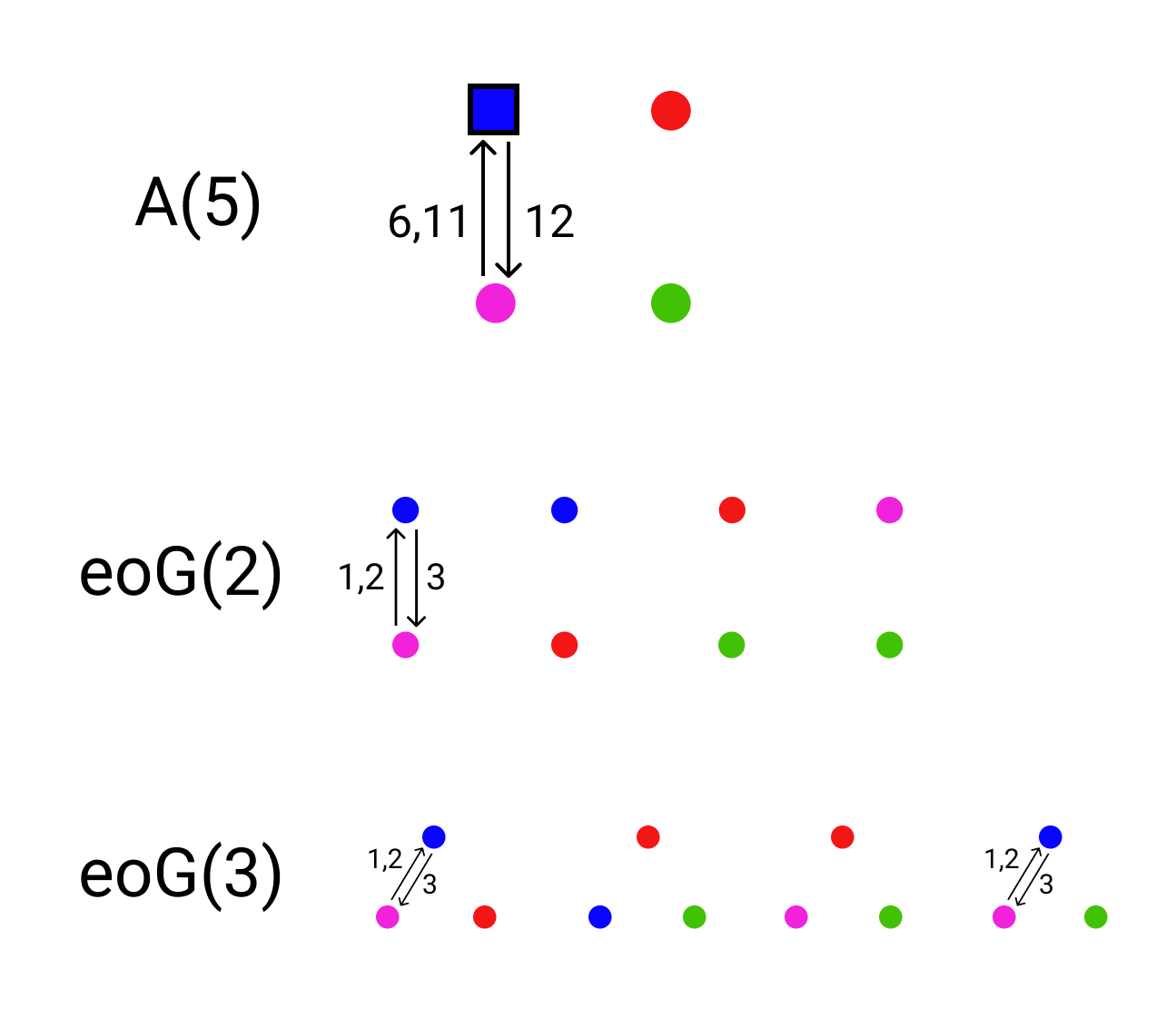}
    \caption{Graphlets of $(M_{\Psi}[A_{5}],p)$.}
    \label{fig:EOGraphlets2}
\end{subfigure}
\caption{The two and three node edge-ordered multidigraphlet decomposition of $M_{\Psi}[A_{4}]$ and $M_{\Psi}[A_{5}]$.}
\label{fig:EOGraphlets Examples}
\end{figure}

In other words, any edge-ordered multidigraph isomorphism  preserves both the topology and the absolute edge-orderings. This notion lead us to  define \textit{edge-ordered multidigraphlets} as equivalence classes of isomorphic edge-ordered multidigraphs where equivalence is with respect to the relative edge-orderings. For instance, consider the edge-ordered multidigraphlets in Figure $\ref{fig:EOGraphlets Examples}$. Note that the graphlets' edges carry a relative ordering, not an absolute ordering.
 
Similar to how we created a free $K$ module $\Lambda$ space for multidigraphlets based on the three parameters $N$, $E$, and $F$, we will characterize a collection of edge-ordered multidigraphlets in a similar way. We can consider the relationship between multidigraphlets and edge-ordered multidigraphlets similar to combinations and permutations. For multidigraphlets, the order does not matter, but for edge-ordered multidigraphlets it does.

\section{Graphlet-based Statistics on Synaptic Signal Graphs} \label{sec:stats}

In this last section we extend a graphlet-based similarity metric in order to compare synaptic signal graphs and the transitions between the them. We focus on the \textit{orbit transition agreement} since it provides a comparison of different temporal networks \cite{Temp Net}. We specifically address the question 'how do the signal dynamics of networks evolve?'. 

Related work is exploring what the topologies are for each of the synaptic signal graphs corresponding to signal propagations between node activations. We will also analyze the frequency of graphlets in order to understand if certain classes of graphlets appear more frequently than others.

\subsection{Orbit transition agreement}
The evolution and time-dependence information of a temporal network is efficiently captured with the orbit transition agreement (OTA) metric, more so than other metrics based on taking frequencies and distributions of graphlets within a network. This is because other metrics do not consider the different transitions of the possible states of orbits within a temporal network. Similarity between two graphs $G_1$ and $G_2$ is given by the average similarity of their graphlet-transition frequency for each graphlet-transition. 

We now extend the orbit transition agreement in \cite{Temp Net} for synaptic signal graphs. After classifying all the graphlets in $G_{\Psi}$, we can  make a matrix. The matrix is a square matrix of size $O_{n}\times O_{n}$, where $O_{n}$ is the total number of orbits generated by all the synaptic signal graphlets of size $n$. The entries $tr_{i,j}$ of this matrix correspond to the number of times an orbit changes from one state to another.
  
\begin{figure}[h]
    \centering
    \includegraphics[width = 6.5in]
   {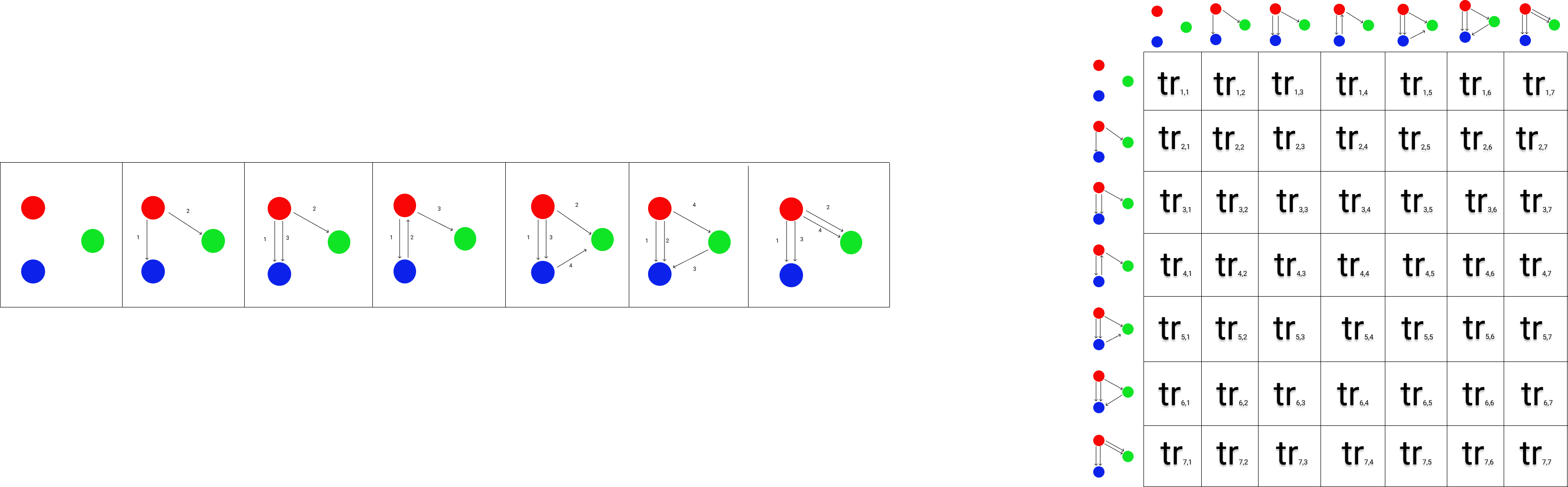}
    \caption{The graphlet-orbit transition matrix corresponding to the possible changes of state of a edge-ordered digraph of size three from one snapshot to another.}
    \label{fig:graphlet_orbit_transitions}
\end{figure}

We then enumerate all graphlet-orbit transitions. The similarity metric is based on an arithmetic mean of orbit-transition differences, which is normalized to reduce biases associated with different sized networks. Normalizing the rows of the orbit transition matrices gives
\begin{center}
\begin{equation}
   ntr_{i,j}=\frac{tr_{i,j}}{\sum_{k=1}^{|O|} tr_{i,k}}
\end{equation}
\end{center}
For any two synaptic signal graphs $G_{\Psi}^{1}$ and $G_{\Psi}^{2}$, we compute their similarity by the average of their graphlet-transition frequency for each graphlet-transition $ntr_{i,j}$. This metric is given by the orbit-transition agreement (OTA):
\begin{center}
\begin{equation}
    OTA(G_{\Psi}^{1}, G_{\Psi}^{2}) = \frac{1}{|O|} \times \sum_{i=1}^{|O|}\sum_{j=1}^{|O|} (1 - |ntr_{i,j}^{G_{\Psi}^{1}} - ntr_{i,j}^{G_{\Psi}^{2}}|)
\end{equation}
\end{center}

\section{Conclusions} \label{sec:conclusions}
We introduced an approach for analyzing the similarity between two dynamic patterns computed by our competitive refractory dynamic model operating on an underlying structural network with a fixed connectivity topology. The technical approach we took was to create directed graphlets with multiple edges to encode multiple signals between connected nodes, and then apply edge-ordering in order to account for variable (synaptic) weight contributions to the running summation of arriving signals at a given target node. These constructions extended graphlets to multidigraphlets and edge-ordered multidigraphlets. A challenge presented by the new graphlet types was the enumeration process. We overcame this by describing the space of graphlets with a vector space whose coordinates represented a multidigraphlet class. 

We studied the graph transitions going from one node activation time to the next. The crucial observation was the fact that there is a subgraph that is preserved between transitions. This creates a constraint on the graphlet-orbit transition matrices. Another important consideration was to conduct pairwise comparisons by only comparing graphlets observed in one or both synaptic signal graphs, which saves computational resources. 

Future work will focus on the algorithms that compute these graphlet-based analyses. In addition, we are examining how edge-graphlets perform in contrast to vertex-graphlets, and we are extending graphlets into a persistent framework, thereby extending the work that has been done with persistent homology and topological data analysis (TDA).

\newpage
\mbox{}

\nomenclature{$G$}{graph}
\nomenclature{$V(G)$}{vertex set}
\nomenclature{$E(G)$}{edge set}
\nomenclature{$G(n)$}{graph of size n} 
\nomenclature{$G_{s}$}{graphlet}
\nomenclature{$\mathcal{G}(n)$}{set of graphlets with size n}
\nomenclature{$O(n)$}{set of all orbits of graphlets with size n}
\nomenclature{$G_{\Psi}$}{synaptic signal graph}
\nomenclature{$G_{U}$}{underlying graph}
\nomenclature{$\leq_{E}$}{total order/precedence on an edge set}
\nomenclature{$\subseteq$}{subset}
\nomenclature{$\rho$}{time a signal becomes traversal}
\nomenclature{$\psi$}{time a signal becomes synaptic}
\nomenclature{$\alpha$}{time a signal becomes terminal}
\nomenclature{$T_{O}$}{observation time}
\nomenclature{$\Psi$}{signal set}
\nomenclature{$(\Psi,\leq)$}{ordered signal set}

\printnomenclature

\appendix
\section{Appendix}
\subsection{Undirected graphlets}

Undirected graphs were first used as a suitable  model for representing any binary relation. Due to its simplicity they were used among different research fields showing a high efficiency for describing the complex structure of what they were modeling. After having being fully used for comparing and classifying networks depending on the frequency and distribution of a small subset of subgraphs, it was discover that similar networks posses a particular type of signature: these are induces subgraph which are called graphlets. Below we show all the possible graphs with their respective orbits up to 5 vertices. Each orbit represents a different topological property.

\begin{figure}[h]  
\centering 
 \includegraphics[width = 10cm, height = 6cm]{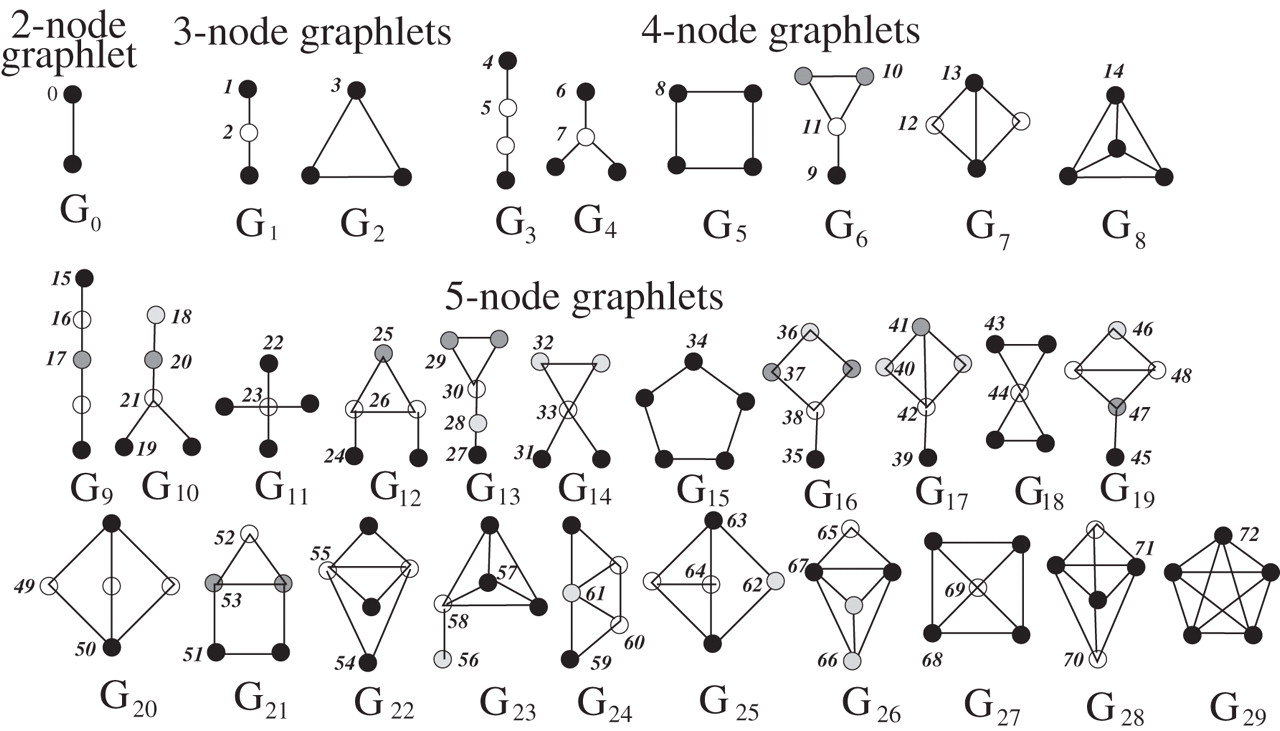}
 \caption{Note: The induced undirected graphs up to five nodes. Reprinted from \cite[Figure 1]{Prz 1}. }
 \label{fig:graphlet_names}
\end{figure}

\subsection{Directed graphlets}

In real-world networks, a directed edge between two nodes represents an interaction of a kind. These interactions can vary from a macroscale to a microscale, reaching social, economical, physical and biological fields: for instance, in biology transcription networks describes all of the regulatory transcription interactions in a cell, whereas in economics one can construct a world trade network to study trade flows. 

Using directed graphs for representing a system allowed us to gather more information about hidden relationships within the system itself. This also led to increased complexity of possible topological states, which were also captured by graphlets. Below we show all the possible graphlets with their respective orbits up to 5 vertices.  

\begin{figure}[h]  
\centering 
 \includegraphics[width = 13cm, height = 8cm]{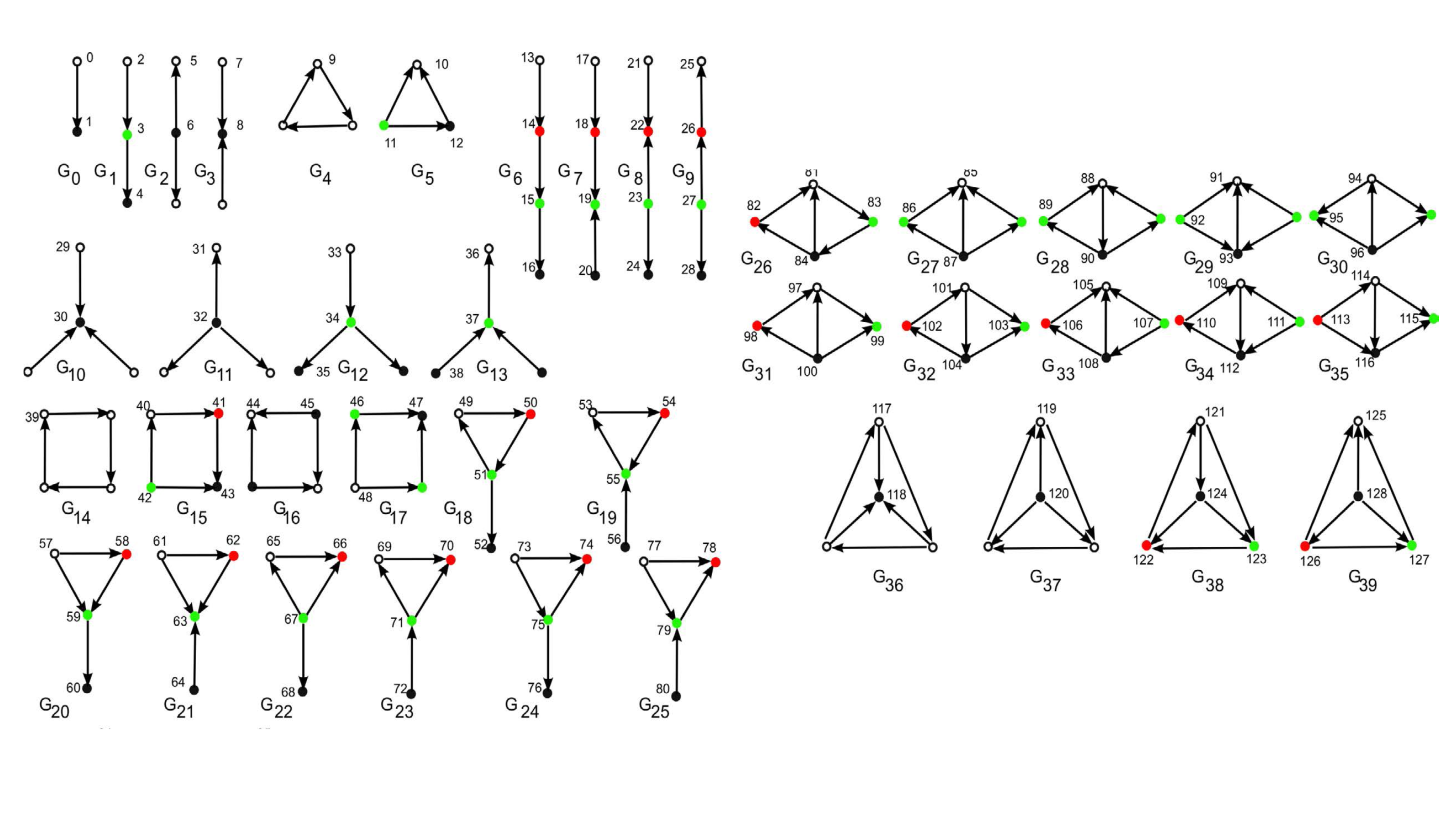}
 \caption{Note: The induced directed graphs up to four nodes. Reprinted from \cite[Figure 1]{Prz 3}. }
 \label{fig:}
\end{figure}

\subsection{Orbit transition matrix, undirected case}

Each orbits within a graphlet carries a particular topological information. These topological features are also induced by the automorphism set, that is, for any automorphism $f$ of the graphlet, any node in a $j$-th orbit goes necessarily to another node which also belongs to the $j$-th orbit. This implies that any graphlet automorphism  induces a permutation of the different orbits sets into themselves. 

When comparing two temporal networks, information is lost with the use of techniques based on the frequency and distribution of graphlets. Neither frequency nor distribution are capable of considering the evolution of the systems. One way to overcome this is taking into account the possible changes of orbit's states. Below we show the graphlet-orbit transition matrix corresponding to the possible orbit transitions of 3-node in the undirected case.

\begin{figure}[h]
    \centering
    \includegraphics[width = 8cm]
   {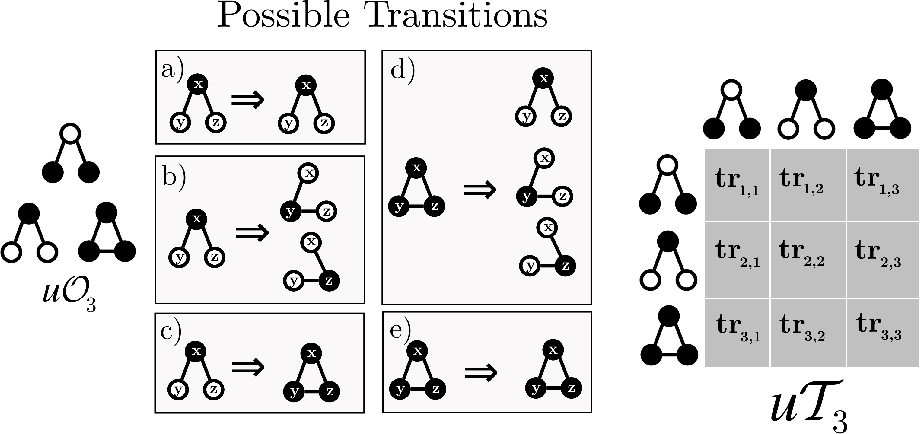}
    \caption{ Each entry $tr_{i,j}$ in the matrix represents the number of times orbit $i$ changes to orbit $j$. Reprinted from \cite[Figure 3]{Temp Net}.}
    \label{}
\end{figure}

\end{document}